\documentclass[preprint]{aastex}

\newcommand{\etal}{et al.\ }
\newcommand{\kms}{\, {\rm km\, s}^{-1}}
\newcommand{\ikms}{(\kms)^{-1}}
\newcommand{\mpc}{\, {\rm Mpc}}
\newcommand{\kpc}{\, {\rm kpc}}
\newcommand{\hmpc}{\, h^{-1} \mpc}
\newcommand{\ihmpc}{(\hmpc)^{-1}}
\newcommand{\hkpc}{\, h^{-1} \kpc}
\newcommand{\lya}{Ly$\alpha$ }
\newcommand{\lyaf}{Ly$\alpha$ forest}

\newcommand{\gmo}{{\gamma-1}}
\newcommand{\bF}{\bar{F}}
\newcommand{\hi}{\mbox{H\,{\scriptsize I}\ }}

\newcommand{\kpa}{k_\parallel}
\newcommand{\vk}{{\mathbf k}}
\newcommand{\df}{\delta_F}
\newcommand{\sF}{{F_s}}
\newcommand{\sdelta}{{\delta_s}}
\newcommand{\seta}{{\eta_s}}
\newcommand{\dt}{\Delta \theta}
\newcommand{\dv}{\Delta v}
\newcommand{\pa}{\parallel}
\newcommand{\pe}{\perp}
\newcommand{\dz}{\Delta z}

\slugcomment{Submitted to ApJ.}

\begin{document}

\title{Toward a Measurement of the Cosmological Geometry at $z\sim 2$:  
Predicting \lya Forest Correlation in Three Dimensions, and the Potential
of Future Data Sets}

\author{Patrick McDonald \altaffilmark{1}}

\altaffiltext{1}{Department of Astronomy, The Ohio State University, 
Columbus, OH 43210; mcdonald@astronomy.ohio-state.edu}

\begin{abstract}

The correlation between
\lya\ absorption in the spectra of quasar pairs can be used to
measure the 
transverse distance scale at $z\sim 2$, which is sensitive to the 
cosmological constant ($\Omega_\Lambda$) or other forms of vacuum 
energy.  Using 
Hydro-PM simulations, I compute the three-dimensional 
power spectrum of the \lyaf\ flux, $P_F(\vk)$, 
from which the redshift-space anisotropy of the correlation
can be obtained.    
I find that box size  $\sim 40 \hmpc$ and resolution $\sim 40 \hkpc$
are necessary for convergence of the calculations to 
$\lesssim 5$\% on all relevant scales, although somewhat poorer resolution 
can be used for large scales.
I compute directly the linear theory bias 
parameters of the \lyaf, potentially allowing simulation results to be 
extended to arbitrarily large scales.  I investigate the dependence
of $P_F(\vk)$ on the primordial power spectrum, the temperature-density 
relation of the
gas, and the mean flux decrement, finding that the redshift-space 
anisotropy is relatively insensitive to these parameters.      
A table of results is provided for different parameter variations. 
I investigate the constraint that can be obtained 
on $\Omega_\Lambda$ using quasars from a large survey. 
Assuming $13 \left(\theta/1'\right)^2$ pairs at separation $<\theta$, 
and including separations $<10'$, a 
measurement to $\lesssim 5$\% can be made if simulations 
can predict the redshift-space 
anisotropy with $\lesssim 5$\% accuracy, or to $\lesssim 10$\% if 
the anisotropy must be measured from the data.
The Sloan Digital Sky Survey (SDSS) will obtain spectra for a 
factor $\sim 5$ fewer pairs 
than this, so followup observations of fainter pair candidates will be 
necessary.
I discuss the requirements on spectral resolution and signal-to-noise
ratio (SDSS quality spectra are sufficient).

\end{abstract}


\section{INTRODUCTION}

The theory of the \lya forest based on gravitational collapse 
of a continuously fluctuating intergalactic medium (IGM) has been 
used to account for the correlation in absorption
along the line of sight to single quasars, and to infer from these
observations the primordial
power spectrum of density perturbations 
(Croft \etal 1999, 2000; McDonald \etal 2000).
Now the calculation should be extended
to correlation across the line of sight, i.e., correlation 
between absorption in spectra of quasars separated by small 
angles in the sky.  
Comparison of theoretical predictions to a measurement
of the full dependence of the correlation function of the transmitted
flux on angle and 
separation will be a valuable test of the \lya forest theory
itself; however, the ultimate purpose of this measurement
is to constrain the cosmological geometry at $z \sim 2$ through the 
Alcock \& Paczy\'nski (1979) test (hereafter, AP test), as 
proposed by McDonald \& Miralda-Escud\'e (1999) and
Hui, Stebbins, \& Burles (1999).  The AP test
is in turn sensitive to the presence of a cosmological constant, 
$\Omega_\Lambda$, or other kinds of vacuum energy.
This paper addresses the need to understand the redshift-space
anisotropy of the \lyaf\ correlation in order to perform an accurate
measurement of cosmological parameters.

We can move aggressively to use the \lyaf\ as a cosmological probe
because a working theory for the formation of the forest has been 
developed in many papers over the 
last decade, using semi-analytic methods and numerical simulations 
(e.g., McGill 1990; Bi 1993; Cen et al. 1994; 
Zhang, Anninos, \& Norman 1995; Petitjean,
M\"{u}cket, \& Kates 1995; Hernquist et al. 1996; Miralda-Escud\'e et al. 
1996; Hui, Gnedin, \& Zhang 1997; Gnedin \& Hui 1998; Theuns et al. 1998).
The theory has been tested by comparing predictions with observed statistics 
of fitted absorption lines and the transmitted flux itself (e.g., 
Bechtold et al. 1994; Dinshaw et al. 1994; Rauch et al. 1997;
Dav\'e et al. 1997; Gnedin 1998;
Crotts \& Fang 1998; Theuns et al. 1999; McDonald et al. 2000;
Zaldarriaga, Seljak, \& Hui 2000).

In this primarily theoretical paper, I present computations 
of the three-dimensional
power spectrum of the transmitted flux (equivalent to the 
correlation function).  
The flux power on large scales is given by the usual redshift-space
formula derived from the linear theory of gravitational 
collapse (Kaiser 1987):
\begin{equation}
P_{F,L}(\vk) = b^2 (1+ \beta \mu^2)^2 P_L(k) ~, 
\label{kaiseq}
\end{equation}
where $\vk$ is the redshift-space wavenumber, $P_L(k)$
is the real-space, linear theory power at $k \equiv |\vk|$, 
$\mu = k_\parallel/k$, $k_\parallel$ is the projection of $\vk$ 
along the line of sight,  
$b$ is a ``bias'' parameter relating flux fluctuations to 
density fluctuations, and $\beta$ is a second parameter describing the 
redshift-space anisotropy.  
I compute the values of both of these parameters from the \lyaf\ theory 
using numerical simulations. [Usually, $b$ and $\beta$ have been discussed 
in the context of galaxy clustering, where 
$\beta \simeq \Omega_m^{0.6}(z)/b$ and $\Omega_m(z)$ is
the matter density (in units of the critical density) at redshift $z$;  
however, for the \lyaf\ $\beta$ is an 
independent parameter (McDonald \etal 2000).]
On small scales, where equation (\ref{kaiseq}) is invalid because of 
non-linear effects, I extract the power spectrum from the 
simulations directly.

A direct computation of the bias parameters is unprecedented.
While $b$ and $\beta$ set the amplitude of the large-scale 
(i.e., linear) power, their values are
in fact determined by the small-scale structure of the field in question 
(i.e., transmitted flux or galaxy density), so the problem of 
computing their values from first principles is generally
non-linear (see Dekel \& Lahav 1998).  In the 
case of the galaxy density, a complete calculation 
is hopeless, because of the importance of star formation to galaxy
formation, although some understanding of the physical nature and 
evolution of galaxy bias has been achieved by combining numerical 
simulations with semi-analytic prescriptions for star formation
(e.g., Blanton et al. 2000; Benson et al. 2000; Cen \& Ostriker 2000; 
Somerville et al. 2001).  However, the Lya forest is
much simpler according to the picture developed in the papers listed
previously. We can simulate the small-scale 
structure in most 
of the volume of the IGM, essentially from first principles, and thus 
compute model predictions for any observable statistic of the \lyaf\ 
transmitted flux, including the large-scale bias!

The observational motivation for this work is the impending flood of quasar
spectra that will need to be analyzed in the near future [e.g., from 
the Sloan Digital Sky Survey (SDSS); York et al. 2000].
If their full potential for measuring 
$\Omega_\Lambda$ through the AP test (and also for measuring other 
parameters through the measurement of the small-scale power spectrum)
is to be exploited, the accuracy of our model predictions
must exceed the level currently used for one-dimensional work 
(e.g., Croft \etal 2000; Zaldarriaga, Hui, \& Tegmark 2000).  
Here I take some first steps toward an accurate
analysis of large data sets that include close pairs and groups of spectra.
I examine some of the modeling uncertainties that have been 
under-studied in recent work focussed on interpreting 
one-dimensional data, including the effect of pressure in 
the simulations, the resolution and box size of the simulations, and 
the detailed dependence of the power spectrum on 
model parameters.  
I attempt to present the results in a form that will 
encourage comparison with other model predictions, and shed some light 
on the issues that need to be considered when planning the massive 
numerical studies that are inevitably needed before the results of a 
precision AP measurement can be believed (this paper can be thought of 
as a pilot study).  
Although I focus on the three-dimensional correlation needed to accomplish
the AP test, many of the issues discussed are also relevant to the
estimation of the mass power spectrum from one-dimensional data.
Finally, I use the computed power spectrum to estimate the potential of
the AP test to constrain $\Omega_\Lambda$ and the requirements on 
signal-to-noise (S/N) and resolution of the data, more realistically than
previously possible.

The plan of the paper is as follows:
In \S 2 I review the basics of the AP test.
In \S 3 I describe and test my procedure for computing the 
power spectrum from simulations given a single set of model parameters. 
In \S 4 I describe the changes in the predicted power when each of the 
model parameters is varied.
In \S 5 I discuss the AP test using my new power spectrum calculations.
The reader who is interested first in cosmology, but not the details of 
the \lyaf\ power spectrum, may want to read sections 3--5 in reverse
order.

\section{MEASURING THE COSMOLOGICAL GEOMETRY USING THE AP TEST}

The function of redshift that relates angular separation ($\dt$) on 
the sky to Hubble flow velocity separation perpendicular to the line 
of sight ($\dv_\pe$) 
can be measured from the correlation function of any observable field
by requiring that the correlation be isotropic in real space.
At high redshift this measurement is sensitive to the 
cosmological constant (Alcock \& Paczy\'nski 1979).
One advantage of this method is the fact
that no assumption of standard candles or rods is required, and it is
therefore independent of evolutionary effects of observed ``objects''.
The total velocity separation between two points along the line of
sight is $\dv_\pa = c \dz /(1+z) = \dv_h + \dv_p$ where $\dz$ is the
redshift separation, $v_h$ is the
Hubble flow velocity and $v_p$ is the peculiar velocity. The 
perpendicular velocity separation is
$\dv_\pe \equiv c f(z) \dt$, where $f(z) = c^{-1} H(z) D_A(z)$, 
$H(z)$ is the Hubble 
constant at $z$, and $D_A(z)$ is the usual angular
diameter distance. With the assumption of isotropy, the real 
space two-point
correlation function of fluctuations, $\xi_r (\dv_h, \dv_\pe)$,
must be a function of $(\dv_h^2+\dv_\pe^2)^{1/2}$ only.
If $\xi_r$ could be measured, it would be a relatively straightforward
matter to measure $f(z)$ by simply demanding isotropy. 
However, generally the large-scale correlations in the universe
are induced by gravitational collapse, and the peculiar velocities make
the correlation function in redshift space anisotropic (Kaiser 1987).
The peculiar velocities introduce 
an anisotropy in the observable (redshift space) correlation function,
$\xi(\dv_p, \dv_\pe)$, of the same order as the difference in $f(z)$ between 
various cosmological models.
In McDonald and Miralda-Escud\'e (1999) 
(see also Hui, Stebbins, \& Burles 1999)
we showed that it is possible to disentangle the effects of geometry and
peculiar velocities and recover the desired global parameters from 
the correlations in the \lya forest absorption.

\begin{figure}
\plotone{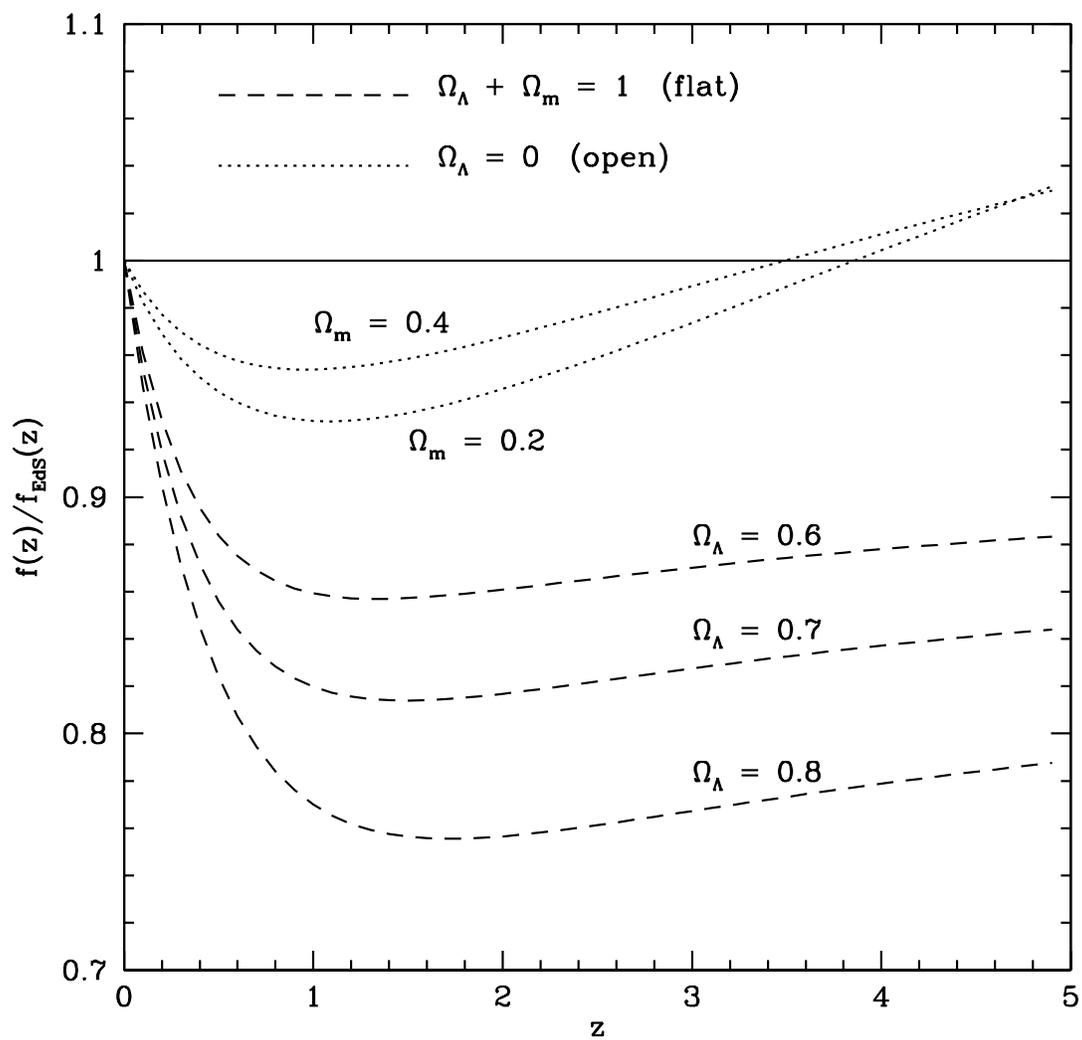}
\caption{Redshift evolution of $f(z)$, relative to an Einstein-de Sitter
model.  The {\it dashed lines} show flat models (with $\omega=-1$), 
while the {\it dotted
lines} show open models.
}
\label{fzev}
\end{figure}

I now show the sensitivity of $f(z)$ to cosmological parameters, assuming
a model containing pressureless matter with density 
$\rho_m = \Omega_m \rho_c$, and a second component with negative pressure,
$p= \omega \rho_\Lambda$, and energy density 
$\rho_\Lambda= \Omega_\Lambda \rho_c$, where $\rho_c$ is the critical 
density.  For simplicity, I assume $\omega$ is a constant.
$f(z)$ is given by
\begin{equation}
f(z)=\frac{E(z)~{\rm S}\left[\sqrt{\Omega_K}\int_0^z (dz'/E(z'))\right]}
       {(1+z)\sqrt{\Omega_K}}~,
\end{equation}
where ${\rm S}(x) \equiv x$ for a flat model ($\Omega_m+\Omega_\Lambda=1$), 
${\rm S}(x) \equiv \sinh x$ for an open 
model ($\Omega_m+\Omega_\Lambda<1$), 
${\rm S}(x) \equiv \sin x$ for a closed model ($\Omega_m+\Omega_\Lambda>1$), 
$\Omega_K=\left|1-\Omega_m-\Omega_\Lambda\right|$, and 
$E(z)=\left[\Omega_m~(1+z)^3+\Omega_K~(1+z)^2+
\Omega_{\Lambda}~(1+z)^{3(1+\omega)}\right]^{1/2}$.
The redshift evolution of $f(z)$ in representative models is shown in 
Figure \ref{fzev} (normalized by 
the value in an Einstein-de Sitter model with $\Omega_m=1$ and 
$\Omega_\Lambda=0$).  This figure shows that the AP test 
is a much more efficient method of measuring $\Omega_m$ in a 
universe with a significant cosmological constant than in an open
universe, with only pressureless matter.  In the flat case, we see that the 
sensitivity of $f$ to $\Omega_\Lambda$
peaks just beyond $z=1$, and is fairly constant with increasing redshift. 

\begin{figure}
\plotone{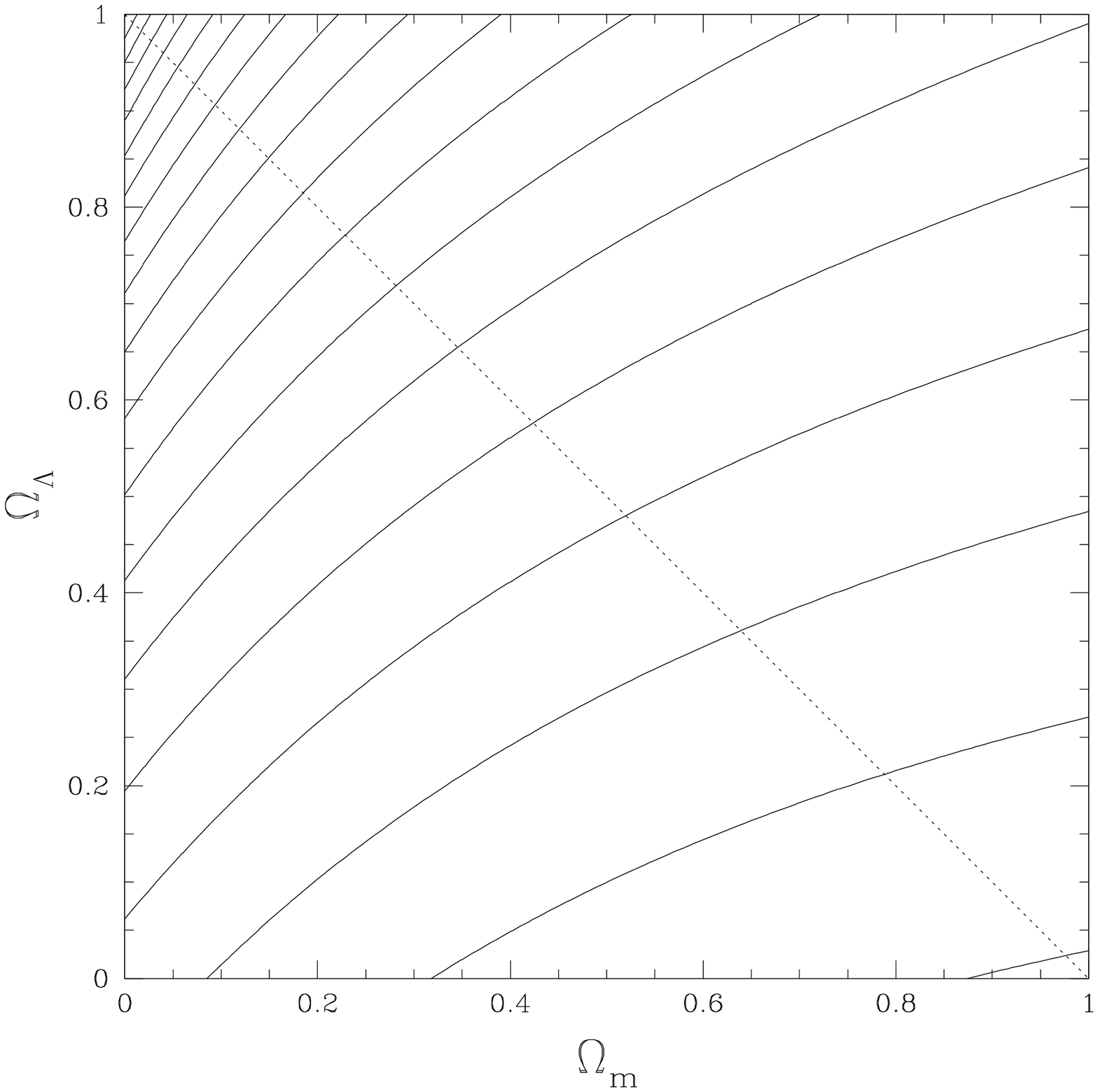}
\caption{{\it solid lines}: contours of constant $f(z)$ (separated by 0.05), 
assuming $\omega=-1$.  {\it dotted line}: $\Omega_m+\Omega_\Lambda=1$.
}
\label{ommomlfcont}
\end{figure}

\begin{figure}
\plotone{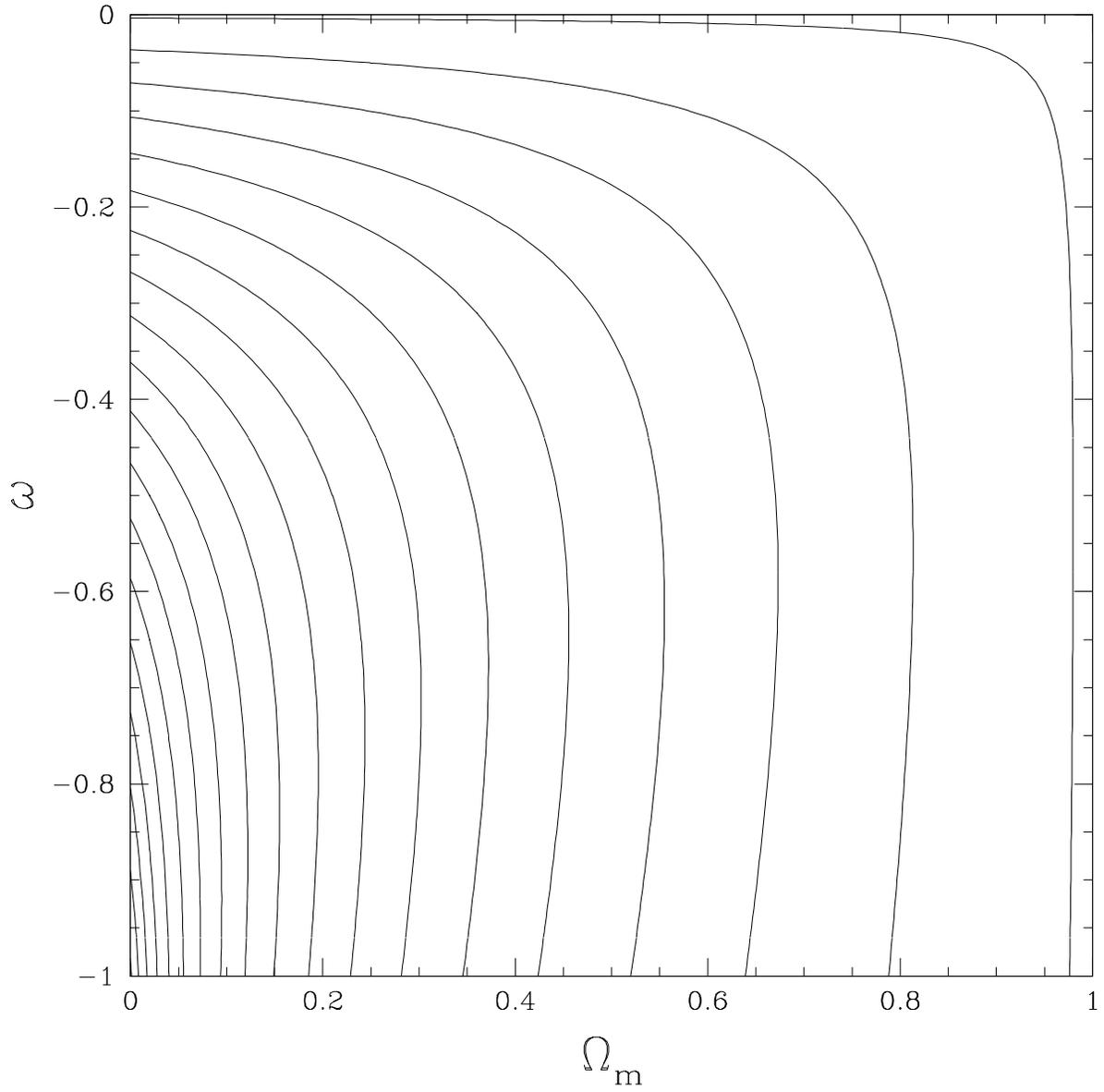}
\caption{Contours of constant $f(z)$ (separated by 0.05), 
assuming a flat universe ($\Omega_m+\Omega_\Lambda=1$).
}
\label{ommwfcont}
\end{figure}

In \S 5 I show that
the effective central redshift for the \lyaf\ AP test using SDSS data
will be $z\simeq 2.25$, so I look more carefully at the sensitivity to 
parameters of $f(z)$ at this redshift.
Figure \ref{ommomlfcont} shows contours of constant $f(z=2.25)$ in the 
$\Omega_m$--$\Omega_\Lambda$ plane, assuming $\omega=-1$.
For a typical model with $\Omega_m \simeq 0.3$ and $\Omega_\Lambda \simeq 0.7$,
the contours are perpendicular to the line that indicates flatness.  
This means that a measurement of $f$ complements the CMB anisotropy 
measurements, which constrain $\Omega_m+\Omega_\Lambda$, and will be a 
cross-check on the type Ia supernova measurements, which have contours 
similar to the \lyaf\ test (recent CMB and SNIa constraints are combined 
in de Bernardis \etal 2000, and Balbi \etal 2000).
Assuming the universe is flat, Figure \ref{ommwfcont} shows contours of 
constant $f$ in the $\Omega_m$--$\omega$ plane.  
The remarkable insensitivity of the \lyaf\ AP test to $\omega$ in models 
with $\omega\lesssim -0.5$ is actually a positive aspect of the test, because 
many of the other future tests 
(e.g., improved CMB, SNIa, and galaxy number counts) have roughly diagonal 
degeneracy contours in the $\Omega_m$--$\omega$ plane (Huterer \& Turner 2000).

Aside from measuring cosmological parameters, simply measuring $f(z)$
at high redshift would test the correctness of currently studied
Friedmann-Robertson-Walker cosmological models in a qualitatively new
regime.    

\section{COMPUTING THE THREE-DIMENSIONAL FLUX POWER SPECTRUM}

The purpose of this section is to establish how I compute the 
three-dimensional flux power spectrum, $P_F(\vk)$, for a given set of 
model parameters, and to explore some of the potential numerical and 
modeling uncertainties.  
In \S 3.1 I describe the simulations that
I use, in \S 3.2 I investigate the effect of 
pressure in the simulations, in \S 3.3 I test the effects of simulation
resolution and box size, in \S 3.4 I introduce a method for extending
the simulation predictions to scales larger than the box size by computing
the linear theory bias parameters for the \lyaf, and in \S 3.5 I
introduce a simple analytic fitting formula that conveniently 
describes the power for all $\vk$.

When I discuss the modeling uncertainties, I will generally define 
better than 10\% accuracy in the ratio of the power along to the power 
across the line of sight to be a ``good'' result.  
The anisotropy of the power is the 
most relevant quantity for the AP test.  In \S 5 I show that 10\% 
accuracy is more than sufficient to interpret existing data,
and data that will exist in the very near future, and is approaching the
accuracy needed for comparison with the full quasar sample of the SDSS.
Furthermore, the approximations made in the Hydro-PM 
simulations that I use (described below) 
are only accurate to $\sim10$\%, so this represents a lower limit on 
the achievable accuracy.  Achieving better accuracy should be a 
straightforward matter of extending the type of study I present
in this paper to include larger and fully hydrodynamic simulations.

For the reasons discussed in Hui \etal (2000), I compute the power in
the fluctuations of $\df({\mathbf x}) \equiv  
F({\mathbf x}) / \bF -1$, where $\bF$ is the mean transmitted
flux, and ${\mathbf x}$ is the redshift-space coordinate (i.e.,
the component of ${\mathbf x}$ along the line of sight is 
$\Delta v_\parallel$, while the components transverse
to the line of sight are described by $\Delta v_\perp$ and an azimuthal
angle).
I use the normalization convention 
\begin{equation}
\left< \df^2 \right> = \int \frac{d^3\vk}
{\left(2 \pi\right)^3} P_F(\vk)~,
\end{equation}
so the flux correlation function is
\begin{equation}
\xi_F\left({\mathbf x}\right)=
\left<\delta_F\left({\mathbf r}\right)
\delta_F\left({\mathbf r}+{\mathbf x}\right)\right>=
\int \frac{d^3\vk}
{\left(2 \pi\right)^3} P_F(\vk)~
\exp\left(-i {\mathbf k} \cdot {\mathbf x}\right)~.
\label{coreqftpow}
\end{equation}
For convenience, I sometimes plot the quantity
\begin{equation}
\Delta^2_F(\vk) \equiv \frac{k^3}{2 \pi^2} P_F(\vk)~,
\end{equation}
where $k\equiv \left| \vk \right|$.
I use $\mu \equiv \kpa / k$ to describe the angle between the wavevector 
$\vk$ and the line of sight, where $\kpa$ is the component
of $\vk$ along the line of sight.

\begin{figure}
\plotone{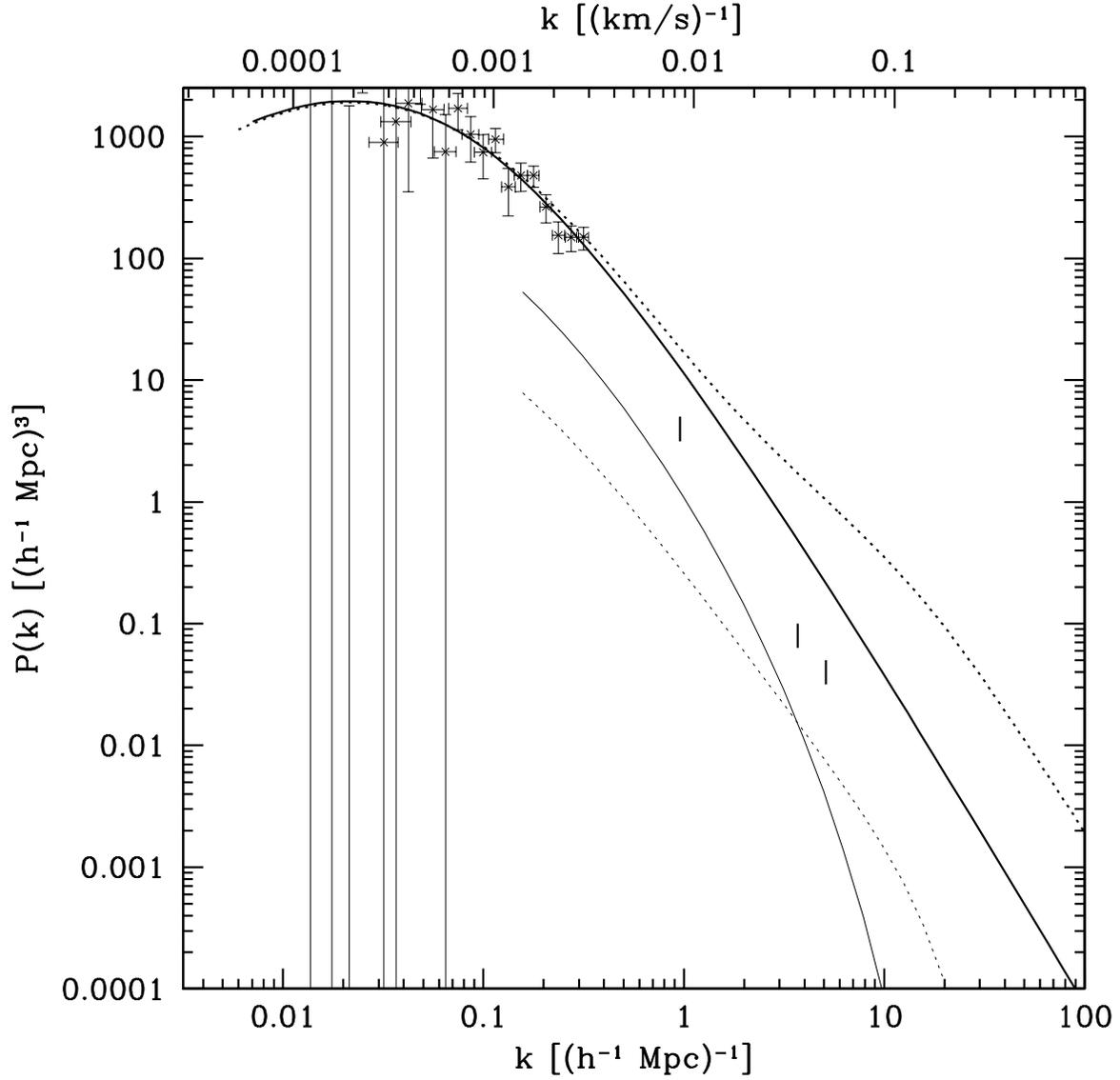}
\caption{\lyaf\ power vs. mass power and galaxy power.
The {\it thin solid} and {\it thin dotted} lines show predictions for
$P_F(\vk)$ along and across the line of sight, respectively, for a 
$\Lambda$CDM model at $z=2$.  
The {\it thick solid} line
shows the linear theory, real space power of the mass fluctuations, while
the {\it thick dotted} line shows the non-linear mass power.
Short, vertical lines indicate the central wavenumbers 
for several recent determinations of the mass power spectrum 
using the one-dimensional flux power (the vertical positions of these
lines are completely arbitrary).  Points with error bars show
the $z=0$ galaxy power spectrum in the linear regime, from Hamilton \& 
Tegmark (2000) (arbitrarily rescaled in amplitude).
}
\label{introfig}
\end{figure}

Before we dive into detailed figures showing the \lyaf\ power
spectrum, it is helpful to see its place in the context of cold dark
matter (CDM) models.
Figure \ref{introfig} compares $P_F(\vk)$ (at $z=2$) to the linear 
and non-linear
power spectra of the mass fluctuations (in real space).  
The linear (thick solid line) and non-linear (thick dotted line) power are 
from the fitting formulas
given by Ma (1998).  The cosmological model is flat $\Lambda$CDM at $z=2$, 
with $\sigma_8=0.79$, $n=0.95$, $\Omega_m=0.4$, and $h=0.65$.  
The flux power is from the analytic fitting formula described
in \S 3.5, with $\mu=1$ for the solid line, and $\mu=0$ for the dotted
line.  The measurements  
by Croft \etal (1999), McDonald \etal (2000), and
Croft \etal (2000) of the mass power from the one-dimensional flux power
were most sensitive to power at the wavenumbers indicated by 
short, vertical lines in Figure \ref{introfig}:  $k=0.96~\ihmpc$, 
$k=5.1~\ihmpc$, and $k=3.7~\ihmpc$, respectively (for my
assumed cosmological model, note that the vertical position of these
lines is {\it meaningless}). 
To show the unique usefulness of the \lyaf\ to constrain small scale 
fluctuations,
I also plot ({\it points with error bars}) the galaxy power spectrum at $z=0$,
in the linear regime, from Hamilton \& Tegmark (2000), with an arbitrary 
rescaling of the amplitude to account for evolution to $z=2$ and bias.  
(the galaxy clustering is measured also on smaller scales, but these results
are difficult to interpret because of our inability to fully simulate 
the formation of galaxies)

Note that throughout 
this paper I plot figures with $k$ measured in $\ihmpc$ on the lower
horizontal axis, and in $(\kms)^{-1}$ on the upper horizontal axis,
transforming between the two using the model in the
simulations; however,
the reader should keep in mind that the
observable coordinates are velocity along the line of sight, and
angular separation transverse to the line of sight.

\subsection{Description of the Numerical Simulations}

In this subsection I discuss
my procedure for computing the flux power spectrum, 
which breaks neatly into two parts:
First, simulations using a given set of cosmological parameters 
(e.g., $\sigma_8$)
and a given thermal history (temperature-density relation at all times)
are evolved to give the baryon density and velocity fields, which are
output at desired redshifts.  Later, using these outputs, simulated spectra 
can be created for different choices of the parameters of the \lyaf\
model (the temperature-density relation at the output time, and the 
normalization of the optical depth). 

\subsubsection{Density and Velocity Fields from HPM Simulations}

For the purpose of this paper, the freedom to run many simulations 
quickly is more important than having the most realistic simulations
possible.
The \lyaf\ is modeled using the Hydro-PM (HPM) approximation to evolve
the baryon density and velocity fields, using the same code described in 
Gnedin \& Hui (1998).  
One set of particles
is used to represent both the baryons and dark matter.  A pseudo-pressure
term, based on an assumed temperature-density relation 
(equation \ref{TDreq} below),
is added to the usual Newtonian force law as follows:
\begin{equation}
\frac{d{\mathbf v}}{dt}+H{\mathbf v} = -{\mathbf \nabla} \phi-
\frac{1}{\rho} {\mathbf \nabla}P \equiv -{\mathbf \nabla}\psi ~,
\end{equation}
with
\begin{equation}
\psi = \phi + \frac{P(\rho)}{\rho}+\int_1^\rho \frac{P(\rho')}{\rho'}
\frac{d\rho'}{\rho'} ~,
\end{equation}
where ${\mathbf v}$ is the particle velocity, $\phi$ is the gravitational
potential, and $P(\rho)\propto \rho^{\gamma}$ is the pressure.

To compute the pressure term, the HPM cod assumes  
a power-law temperature-density relation, 
\begin{equation}
T=T_0 \Delta^\gmo ~,
\label{TDreq}
\end{equation}
where $\Delta$ is the density in units of the mean density.
This form arises naturally for gas expanding adiabatically
in ionization equilibrium with a background radiation field, where the
temperature is set by photoionization heating and adiabatic cooling 
(Hui \& Gnedin 1997).  Fully hydrodynamic simulations confirm that
equation (\ref{TDreq}) provides a reasonable approximate description of the
temperature-density relation for $\Delta \lesssim 5-10$, above which shock
heating leads to substantial dispersion in the temperatures at fixed 
densities (Croft et al. 1997; Theuns et al. 1998).  
Since most \lyaf\ absorption is by gas with 
$\Delta < 5$, it is not too surprising that the HPM simulation method
can give good agreement with the results of fully hydrodynamic simulations 
(Gnedin \& Hui 1998).  
In practice, I usually use
the parameter $T_{1.4} \equiv T_0~1.4^{\gmo}$ in 
place of $T_0$, because McDonald \etal (2001) measured the temperature
most precisely at
$\Delta=1.4$.

For the purpose of computing this pressure term only, the parameters 
of the temperature-density
relation in my simulations are set to the measured values 
from McDonald \etal (2001):
$T_{1.4}$=(20100, 20300, 20700) K, and $\gmo=$(0.43, 0.29, 0.52),
at $\bar{z}=$(3.9, 3.0, 2.4), with reionization assumed to take place
suddenly at $z=7$, when the gas is raised from zero temperature to a 
constant temperature
$T=25000$ K (i.e., $\gmo=0$).  I interpolate linearly between the given 
redshifts.  As I discuss further below,
within the bounds allowed by the current 
observational measurements of the temperature-density relation, the 
detailed thermal history used to evolve
the simulation is not important to our power spectrum results, at the 
level of accuracy that we are interested in this paper.  For this 
reason, when I vary the temperature-density relation used to compute
the recombination coefficient and thermal broadening when creating 
spectra (see \S 3.1.2), I do not need to re-run the 
simulations with a different
thermal history.

The mesh cells in my simulations are always
equal in size to the spacing between particles. 
Interpolation from the particles to the mesh is performed using 
triangular-shaped clouds (hereafter TSC; Hockney \& Eastwood 1988).
I will refer to the length of the box
along an axis as $L$, and the number of particles as $N$.

\subsubsection{Simulated Spectra}

After the evolution of a simulation is completed, I create simulated 
spectra from which
I measure the power spectrum.  
Entire simulation boxes are converted into spectra by using each grid
cell in a face of the box as the origin
for a line of sight through the simulation cube, i.e., 
a spectral pixel is generated
for every cell in the box (in practice, I under-sample the transverse
directions by factors of 2 because this saves computer time and does not 
introduce any noticeable error).  The power spectrum is then obtained 
by a three-dimensional FFT.

The first step in creating spectra is to convert the HPM particle positions 
and velocities into density and velocity fields on a grid.  This is done
using the same TSC interpolation and grid spacing that was used in the 
evolution of the simulation.  

Next, the baryon density is converted to \hi density, $n_{HI}$, by 
assuming the gas is
in ionization equilibrium with a uniform ionizing background, i.e., 
$n_{HI} \propto T^{-0.7} \Delta^2 / \Gamma$, where $\Gamma$
is the ionization rate, and $T$ is the temperature of the gas 
[I use $\alpha(T) \propto T^{-0.7}$ for the recombination coefficient].  
Assuming the temperature is given by equation (\ref{TDreq}), I obtain 
from $n_{HI}$ the optical depth in each cell in real-space (i.e., 
before the effects 
of peculiar velocities and thermal broadening),
\begin{equation}
\tau_R = \tau_0(z)~\Delta^{\beta} = 1.41
\frac{\left(1+z\right)^6~ \left(\Omega_B h^2\right)^2}
{T_4^{0.7}~ h E(z)~ \Gamma_{-12}(z)}~\Delta^{\beta}~,
\label{taueq}
\end{equation}
where $E(z)=H(z)/H_0 \simeq \Omega_m^{1/2}~
(1+z)^{3/2}$ (I use the exact version), $H(z)$ is the Hubble 
constant at redshift $z$, $H_0$ is the
present Hubble constant, $h=H_0/(100 \kms \mpc^{-1})$, $\Omega_B$ is the
baryon density in units of the critical density (at the present time),
$T_4=T_0/(10000~{\rm K})$,
$\Gamma_{-12}(z)$ is the photoionization rate of hydrogen in units of
$10^{-12}\, {\rm s}^{-1}$, and $\beta = 2-0.7~(\gmo)$.

Finally, to construct the spectrum along each line of sight, I 
evaluate the integral 
\begin{equation}
\tau(x) = \int \tau_R(x') ~
W\left[x-x'-v_\parallel\left(x'\right), T\left(x'\right)\right] dx' ~,
\end{equation}
where $x$ and $x'$ are periodic velocity coordinates labeling the cells
in the simulation in real-space and redshift-space respectively, $T(x)$
and $v_\parallel(x)$ are respectively the temperature and the velocity 
along the line of sight of the gas at $x$, 
\begin{equation}
W\left(x, T\right)=
\exp\left[-\frac{1}{2}\frac{x^2}{\sigma^2(T)}\right]/
\left[2 \pi \sigma^2(T)\right]^{1/2}~,
\end{equation}
and $\sigma(T)=9.1 \kms \left(T/10000~{\rm K}\right)^{1/2}$.
In practice, the details of the numerical implementation of this integral
may make some difference to my results (specifically the resolution test).
I use redshift-space pixels identical in size to the real-space cells in 
the simulation.  I account for the expansion or contraction of cells by
translating each cell-edge in real-space into 
redshift-space using the average velocity of the cells that the edge
separates. The optical depth contributed by each real-space cell is 
distributed to multiple redshift-space pixels
based on its fractional overlap with each.  The different contributions
to a redshift-space pixel are thermally broadened separately, based on
the temperature of the originating real-space cells.
The observable quantity is $F(x)=\exp\left[-\tau\left(x\right)\right]$.

Because the values of the parameters that combine to form $\tau_0$ are 
uncertain, especially the value of $\Gamma_{-12}$, $\tau_0$ is
effectively an unknown parameter.
The mean transmitted flux, 
$\bF$, is more directly observable, so I use it as the independent 
parameter, determining $\tau_0$ 
by requiring that $\bF$ in the simulation has the specified value.
Unless otherwise indicated, all comparisons between simulations are
at fixed $\bF$.  Note that many of my conclusions may be sensitive
to large changes in the value of $\bF$ 
(because $\bF$ sets the typical density 
to which the power spectrum is sensitive), so they should not be assumed 
to hold at redshifts where the mean flux is much higher or lower.

In summary, the \lyaf\ in my model, for a given density and velocity
field (determined by the cosmological model), is 
specified by three free parameters:  $\bF$, $T_{1.4}$, and $\gmo$.  

For the numerical tests in \S 3.2 and \S 3.3 I use a set of
simulations originally created for comparison with the
hydrodynamic simulation L10 in Miralda-Escud\'e \etal (1996).
The outputs are at $z=2$, with cosmological parameters 
$\sigma_8=0.79$, $n=0.95$, $\Omega_m=0.4$, 
$\Omega_\Lambda=0.6$, $\Omega_B=0.0355$, and $h=0.65$, and
\lyaf\ parameters $\bF=0.818$, $T_{1.4}=15517$ K, $\gmo=0.49$.
Starting with the box size test at the end of \S 3.3, 
and throughout the rest of the paper, I use a different set
of simulations, output at $z=2.25$, and change the \lyaf\ model 
parameters to
$\bF=0.8$, $T_{1.4}=20000$ K, and $\gmo=0.5$ [consistent with the
temperature measurement in McDonald \etal (2001)].  
I show in \S 5 that $z \simeq 2.25$ is
approximately the redshift where the most data will be available to 
perform the AP test.  

Throughout this paper I actually
included deviations from equation (\ref{TDreq})
when I computed the recombination coefficient and thermal broadening
for spectra,
in order to facilitate a future comparison between HPM results and 
simulations like the
fully hydrodynamic simulation referred to as L10 in 
Miralda-Escud\'e \etal (1996). 
I add to the specified power-law the deviation from a power-law
found in that simulation,
i.e., the temperature that I use at a point with overdensity $\Delta$ is
$T=T_0 \Delta^\gmo + \delta T(\Delta)$, where $\delta T(\Delta)$ was
obtained from the hydro simulation (see McDonald \etal 2001, Figure 1).
This additional term makes less than 6\% difference in the flux power 
spectrum for $k<10~(\hmpc)^{-1}$ [and $<4$\% for $k<4~(\hmpc)^{-1}$].
To promote reproducibility, the values of $\delta T(\Delta)$ that I
used are available on request.

\subsection{HPM vs. PM}

Ideally, we would always use fully hydrodynamic simulations to model
the \lyaf, like the one 
used to simulate the one-dimensional power spectrum 
in McDonald \etal (2000).    
Unfortunately, these simulations are very expensive to run, leading 
Croft \etal (1999, 2000) 
and Zaldarriaga \etal (2000) to use simple particle-mesh (PM) simulations.
In this paper, except for in part of this subsection, I use the 
intermediate HPM method
[see McDonald \& Miralda-Escud\'e (2001), Meiksin \& White (2000), and
Ricotti, Gnedin, \& Shull (2000) for some tests and applications].
The only difference between the HPM and PM simulations is the former 
include a simple calculation of pressure, based on an assumed 
temperature-density relation, as described above.  The primary 
difference between the
HPM and full-hydro techniques is the latter include a full calculation
of the temperature evolution of every element of gas and they include 
shocks.  The most important practical consequence of this difference is 
probably the more realistic values for the temperatures used to compute
the recombination coefficient and thermal broadening for the gas.

Before I make detailed comparisons between power spectra 
under different modeling assumptions,
I make one relevant observation about the realities of \lyaf\ data analysis:  
McDonald \etal (2000) showed that $P_{F,1D}(k)$ is significantly influenced
by the presence of metal lines if $k \gtrsim 0.1 (\kms)^{-1}$.  This
means that any measurement of \lyaf\ statistics that is sensitive to
power on this scale is suspect.  Therefore, when I discuss the 
simulation predictions, I will not be very concerned about the 
power in the simulations 
at $k \gtrsim 0.1~(\kms)^{-1}$ [$k\gtrsim 10~\ihmpc$].

\begin{figure}
\plotone{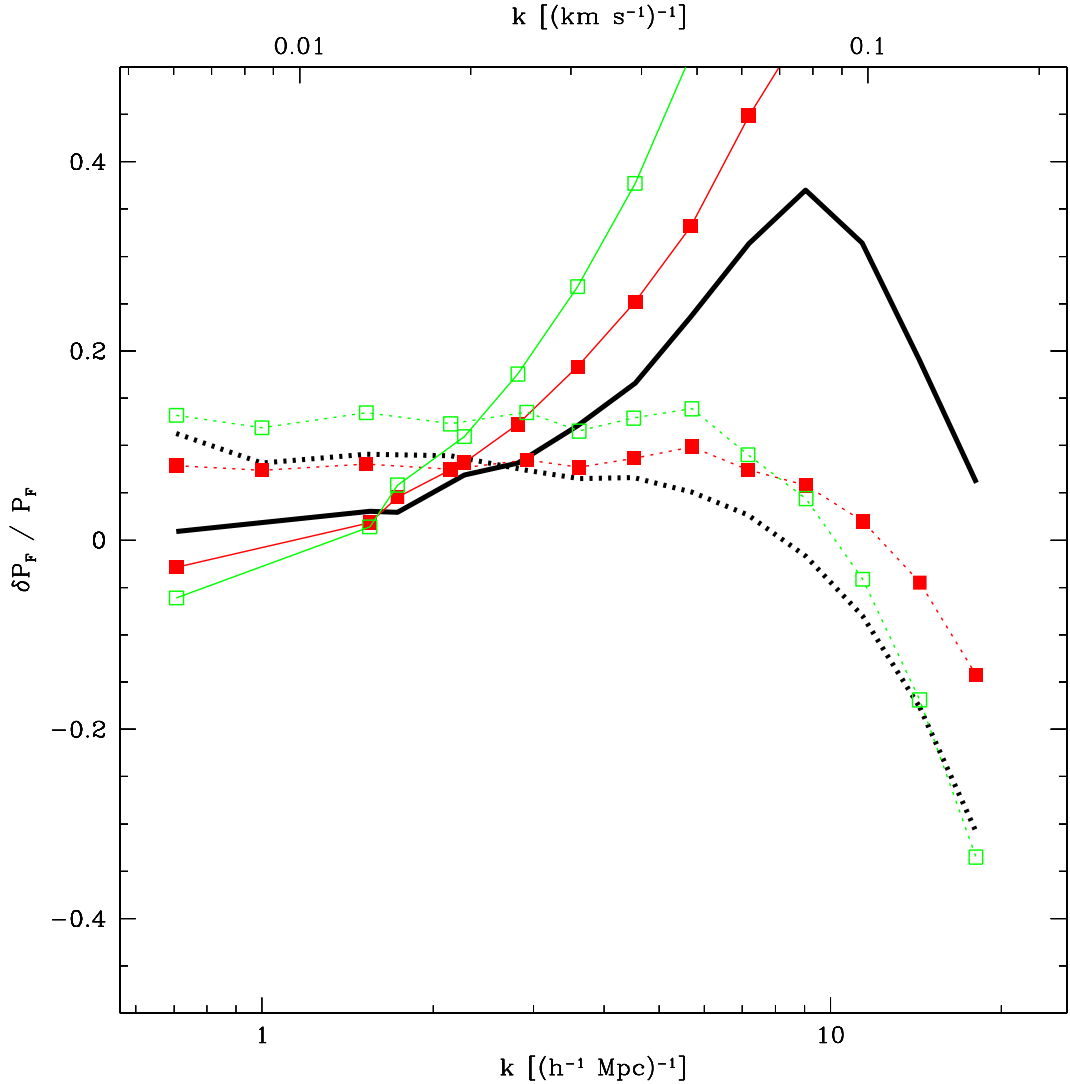}
\caption{Fractional changes in power using different pressure 
approximation methods.
All curves are relative to a simple PM simulation, and in all cases the
{\it solid lines} show power along the
line of sight ($0.75<\mu<1.0$) and the {\it dotted lines} show power across 
the line of sight ($0.0<\mu<0.25$).
The {\it thick, black lines} show the difference between HPM and PM power. 
The squares connected by thin lines show 
Gaussian smoothing, $\exp[-(k r)^2/2]$, applied to the mass distribution,
with $r=35 \hkpc$ ({\it red with filled squares}) and $r=49 \hkpc$ 
({\it green with open squares}).
}
\label{preseff}
\end{figure}

We expect that the pressure in an HPM simulation will smooth the gas
on small scales, relative to a PM simulation. For example, in linear 
theory, and assuming temperature evolution $T \propto (1+z)$, the 
suppression factor for a Fourier mode, $\delta(k$), with wavenumber $k$ 
can be 
calculated analytically:  $\delta(k)/ \delta_0(k)=1/[1+(k/k_J)^2]$, 
where $\delta_0$ is the amplitude the mode would have in absence of 
pressure, $k_J=a \sqrt{4 \pi G \rho}/c_s$,
$a$ is the expansion factor and
$c_s$ is the sound speed of the gas (e.g., Gnedin \& Hui 1998).  
Because the \lyaf\ structure is not linear, and the thermal 
evolution is different from $T \propto (1+z)$, the above formula 
will not hold quantitatively in my simulations, but we should see the 
same general effect.
In Figure \ref{preseff} I compare the flux power spectra from HPM and PM
simulations (with identical initial conditions, $L=8.89\hmpc$, and 
$N=256^3$), plotting the ratio of
HPM to PM power as the thick, black line.  
Throughout this paper (except in Figure \ref{introfig}), 
power measured from simulations
``along the line of sight'' refers to a bin where $0.75<\mu<1.0$, while
power ``across the line of sight'' refers to $0.0<\mu<0.25$.
We see the expected
suppression by pressure of small-scale power 
transverse to the line of sight (for reference, the ratio of HPM to PM power 
is well fit by $1.1 \exp[-(k r)^2]$, where $r=37\hkpc$),
but an increase in the power along the line of sight.  
This increase,
which occurs at $k$ values where the power is already strongly suppressed by 
non-linear peculiar velocities, is probably the result of suppression of
small-scale power in the velocity field.  
Of course, the change in $P_F(\vk)$ caused by pressure differences for 
reasonable changes in the temperature near the measured value 
(see McDonald et al. 2001) is 
much smaller than the difference between PM and HPM simulations.  
For a 4000 K change in temperature, the change in $P_F(\vk)$ is less
than 2\% at $k< 8 ~\ihmpc$, justifying the assumption I make later
that the only significant effect of changing the temperature is to 
change the thermal broadening and recombination coefficient in the 
creation of spectra.  

The effect of pressure on the power spectrum can, in principle, extend to 
all scales;
however, we see that the influence of pressure on large-scale power is 
rather small.  This bodes
well for the general idea that we can substitute HPM approximations
for fully hydrodynamic calculations, at least for measurements at relatively
large separations.
That is, if the difference in 
power between PM and HPM is only 10\%, it seems unlikely that
any inaccuracy in the HPM pressure relative to full-hydro 
will be significant  
(recall that $\sim 10$\% accuracy in the ratio of power 
along to across the line of sight is roughly the level
I am aiming for in this paper).
The main outstanding difference
between a full-hydro and an HPM simulation is the scatter in temperature
at a single density, especially at high densities.  
A detailed test of the systematic differences between 
the two will be the subject of future work.

In their paper interpreting the observed one-dimensional flux power spectrum
results of McDonald \etal (2000), 
Zaldarriaga \etal (2000) attempted to mimic the effect of pressure forces
by smoothing the dark matter density in PM simulations before creating
spectra.  The lines highlighted by squares in Figure \ref{preseff} show
the results of this method for the three-dimensional power, 
where a Gaussian smoothing, 
$\exp[-(k r)^2/2]$, was applied to the mass density and momentum fields, 
using $r=35\hkpc$ (red line with filled squares) and $r=49\hkpc$ (green
with open squares).  While the results have a similar trend 
as the results from the HPM approximation, the relative strength of the effects
in different regions of $\vk$-space are quantitatively different.
Note also that Zaldarriaga \etal (2000) allowed $r$ to be as high 
as $\sim 280 \hkpc$ --
clearly much too high to give even a rough representation of pressure for
a reasonable thermal history of the gas. 

\subsection{Simulation Resolution and Box Size}

To make accurate power spectrum predictions, it is necessary to be
sure we sufficiently resolve the \lyaf\ structure.  As we saw in the
case of pressure, resolving small-scale structure is not only required
to correctly predict the small-scale power, but also to 
predict the large-scale  power (i.e., recall the factor of 1.1 increase,
caused by pressure,
in the large-scale power across the line of sight in Figure \ref{preseff}).
This is analogous to the need to
correctly simulate the formation of individual galaxies before galaxy 
bias can be predicted.  
I investigate the resolution effects by running matched sets of 
simulations with identical box sizes but differing numbers of particles.
The initial mode amplitudes in the different resolution simulations are 
identical up to the Nyquist frequency of each grid. 

\begin{figure}
\plotone{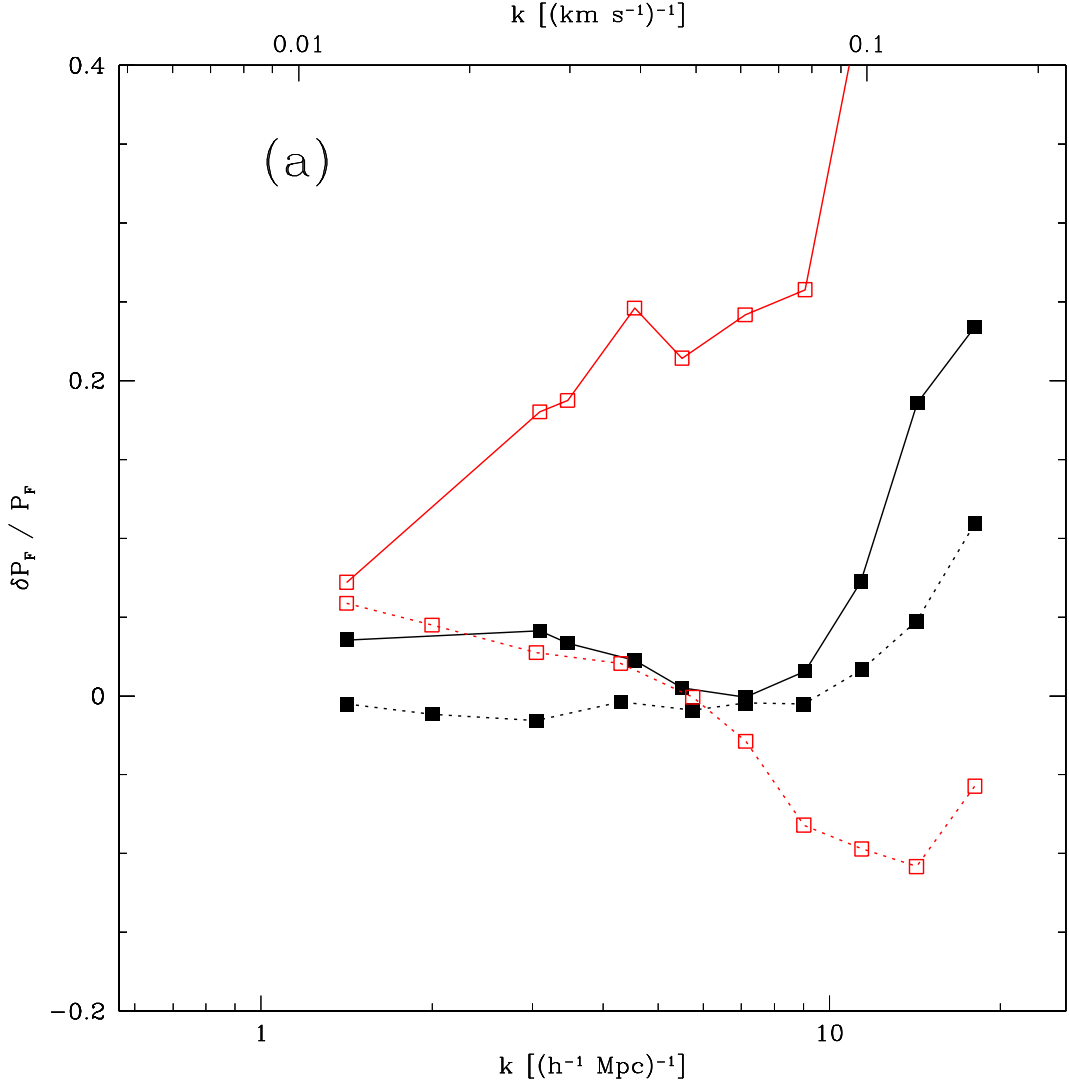}
\caption{Resolution test in an $L=4.44 \hmpc$ box (a) and 
an $L=8.89 \hmpc$
box (b).  Black lines highlighted by filled squares show the ratio 
(minus 1) of
power for $N=128^3$ particles to power for $N=256^3$ particles, along 
({\it solid line}, $0.75<\mu<1.0$), and across 
({\it dotted line}, $0.0<\mu<0.25$) the line of sight.
Red lines with open squares show the ratio of $N=64^3$ to $N=256^3$.   
}
\label{restest}
\end{figure}

In Figure \ref{restest}(a) I compare the power 
for $N=128^3$ ({\it solid line}), 
or $N=64^3$ ({\it dotted line}) to the power for $N=256^3$, using 
a box size 4.44 $\hmpc$. 
The quantity plotted in the figure is 
$\delta P / P \equiv \left(P_N - P_{256}\right)/P_{256}$.
\begin{figure}
\plotone{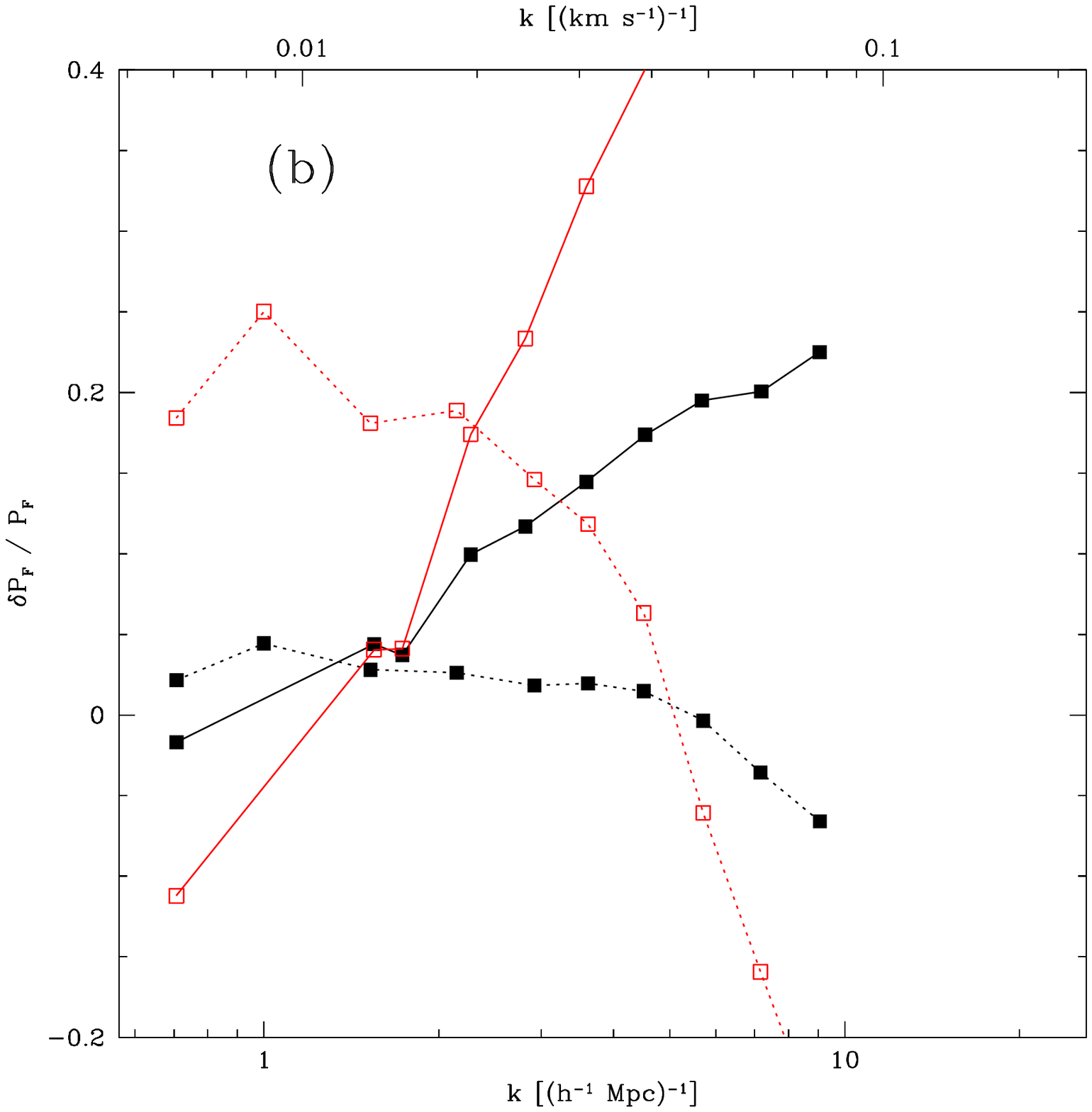}
\end{figure}
The $256^3$ and $128^3$ simulations agree very well [better
than 5\% for $k<10~\ihmpc$] but the
$64^3$ box shows larger disagreement.  (Note that the effect of
insufficient resolution is qualitatively similar to the effect
of pressure seen in Figure \ref{preseff}, and that the required
resolution for convergence is roughly the scale of 
the smoothing by pressure.)  
I conclude that, for a mean particle spacing of $\sim 35 \hkpc$,
the \lyaf\ structure is 
effectively resolved, and that a factor of 2 or so worse resolution
can be used if a high level of accuracy is not required, or if
the sensitivity of the
measurement is carefully restricted to large scales.  
I test this conclusion further in 
Figure \ref{restest}(b), where I use an 8.89 $\hmpc$ box.  
The results are similar, but we see that decreasing the resolution
quickly becomes quite harmful.  These results are not improved 
if I compare simulations at fixed optical depth 
normalization ($\tau_0$) instead of fixed $\bF$.

\begin{figure}
\plotone{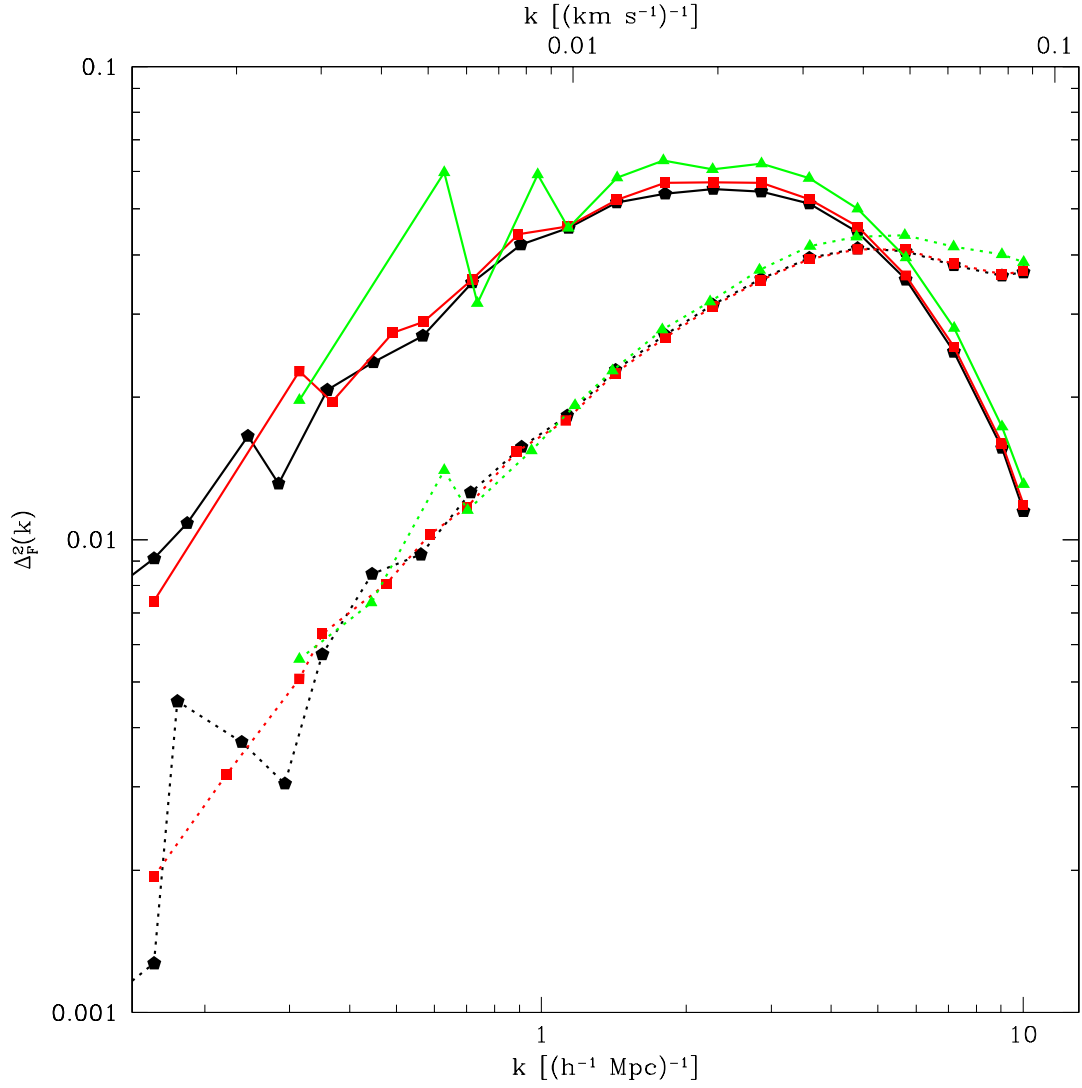}
\caption{Box size test.  Black, red, and green lines (highlighted by
{\it pentagons}, {\it squares}, and {\it triangles}) show the 
power in $L=80$,
40, and 20 $\hmpc$ boxes, with $N=512^3$, $256^3$, and $128^3$
particles, respectively (i.e., identical resolution).  
Power along (across) the line of sight is
indicated by {\it solid} ({\it dotted}) lines.
}
\label{boxsize}
\end{figure}

In addition to obviously limiting the scale on which predictions can be made,
the limited size of the simulation boxes can also influence the power 
spectrum results on all scales, because of the non-linear coupling of modes 
during gravitational evolution.
In Figure \ref{boxsize} I compare 80, 40, and 20 $\hmpc$ boxes, 
with $512^3$, 
$256^3$ and $128^3$ particles, respectively (it is not convenient
to plot ratios because the $k$-modes are not identically spaced 
for different sized boxes). 
Power along (solid lines) and across (dotted lines) the line of sight 
refers to $0.75<\mu<1.0$ and $0.0<\mu<0.25$, as usual. 
For the lowest $k$ modes in the box, it is difficult to see the 
precise level of agreement, because of statistical errors and 
discreteness in the values of $\vk$ represented.  I 
return to the issue of large scale convergence in the 
next subsection, where my computations of the linear theory
bias are much less sensitive to statistical uncertainty.  
At higher $k$ there is a systematic decrease
in power as the box size is increased (the error for a 10 $\hmpc$ box is
even larger than those shown).
This is not surprising because the mass fluctuations are not
perfectly linear even on the scale of the 40 $\hmpc$ box (see Figure
\ref{introfig}).
Fortunately, the difference between the 80 and 40 $\hmpc$ boxes is small 
enough that we can consider the 40 $\hmpc$ box to be sufficiently large. 

My conclusion in this subsection is that 40 $\hmpc$ simulations with
$\sim 1024^3$ particles should be sufficient to {\it completely 
simulate the \lyaf\ flux power spectrum}, to $\sim 5$\% 
accuracy; however, simulations of this size are beyond the scope of this
work.  In the rest of the paper, I solve the problem of competing demands
for large box size and high resolution by splicing together the power 
spectra from pairs of large and
small box simulations, using $L=40 \hmpc$ for the large-scale power, and 
$L=10 \hmpc$ for the small-scale power, both with $N=256^3$.  

In words (equations will come later), I correct for the poor resolution of 
the $L=40\hmpc$, $N=256^3$ boxes by comparing 
$L=10\hmpc$, $N=64^3$ simulations, which have the same resolution, to
$L=10\hmpc$, $N=256^3$ simulations, which have sufficient resolution.
For $k > 2 \pi / (10 \hmpc)$ (the minimum $k$ present in an $L=10\hmpc$
box), I can correct the power at a given $k$ in the $L=40 \hmpc$ simulation 
by the ratio, at the same $k$, of the 
power in the $L=10\hmpc$, $N=256^3$ simulations to the 
power in the $L=10\hmpc$, $N=64^3$ simulations.  For larger scales, 
$k< 2 \pi / (10 \hmpc)$,
I assume (to be checked below) that I can use the $k=2 \pi / (10 \hmpc)$
correction factor, i.e., that the correction factor is independent of $k$.
For small scales, $k\gtrsim 64~\pi/(40 \hmpc)$ (one fourth of the 
Nyquist wavenumber of the large box) the resolution correction 
ceases to be a small factor (see Fig. \ref{restest}), so at this $k$
I switch from using the (corrected) power from the $L=40\hmpc$ box
to using the power from the $L=10\hmpc$, $N=256^3$ simulation.  
I correct for the limited box size by a $k$-independent factor which 
is the ratio, at the splice point $k = 64~\pi/(40 \hmpc)$, of the 
power in the $L=40\hmpc$, $N=256^3$ simulations to the power in 
$L=10\hmpc$, $N=64^3$ simulations.  

In equations, the method just described for computing the corrected power, 
$P_F'(\vk)$, is the following:
For $k_{{\rm min},10} < k < k_{{\rm Nyq}, 40}/4$ 
[where $k_{{\rm min}, 10} = 2 \pi / (10 \hmpc)$ and
$k_{{\rm Nyq}, 40} = 256~\pi /(40 \hmpc)$], I use the formula 
\begin{equation}  
P_F'(\vk)\equiv P_{F,40,256}(\vk) 
\frac{P_{F,10,256}(\vk)}{P_{F,10,64}(\vk)}~,
\end{equation}
where $P_{F,L,N^{1/3}}(\vk)$ is the flux power in simulations with 
box size $L$ and number of particles $N$.
For $k<k_{{\rm min},10}$, I use
\begin{equation}  
P_F'(\vk)\equiv P_{F,40,256}(\vk) 
\frac{P_{F,10,256}(k_{{\rm min},10}, \mu)}
{P_{F,10,64}(k_{{\rm min},10}, \mu)}~,
\label{lowkcor}
\end{equation}
and for $k>k_{{\rm Nyq}, 40}/4$ I use
\begin{equation}  
P_F'(\vk)\equiv P_{F,10,256}(\vk) 
\frac{P_{F,40,256}(k_{{\rm Nyq},40}/4, \mu)}
{P_{F,10,64}(k_{{\rm Nyq},40}/4, \mu)}~.
\end{equation}
In practice, these correction
factors are interpolated between bins in $k$ and $\mu$ 
(because the $\vk$ spacing
is different in simulations with different $L$).
Ultimately, more exact results will be obtained by running larger 
simulations, but, judging from the preceding 
tests, this procedure should
give reasonably accurate results.  

\begin{figure}
\plotone{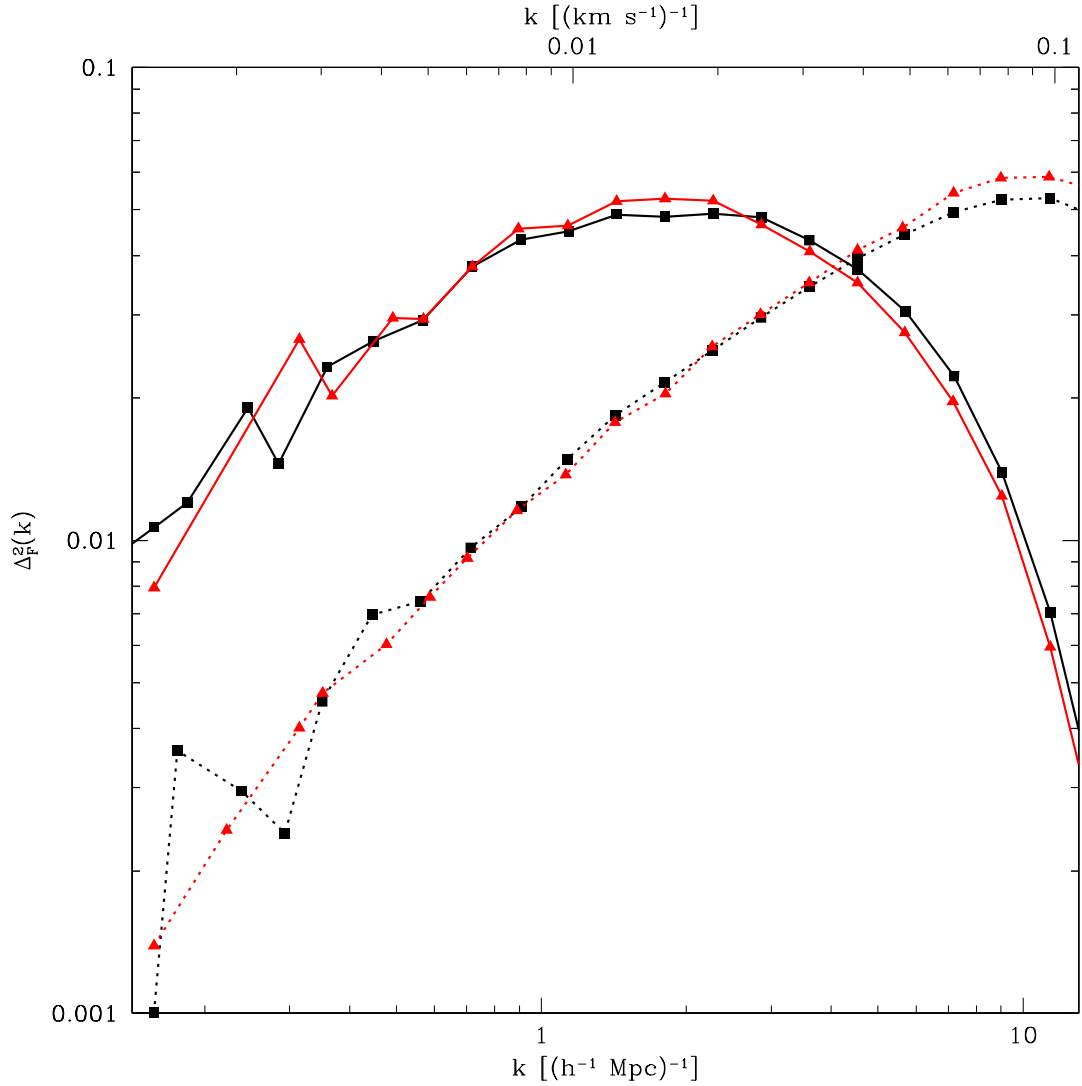}
\caption{Test of the corrections for box size and resolution.  Solid lines
show power along the line of sight, dotted lines across the line of sight.
Red lines (highlighted by triangles) show the results of our standard 
combination of $L=40 \hmpc$ and $L=10 \hmpc$ boxes with $N=256^3$.
Black lines (squares) show the combination of an $L=80 \hmpc$ and 
a $L=20\hmpc$ 
box, with $N=512^3$.  
}
\label{fullprocconv}
\end{figure}

Figure \ref{fullprocconv} shows a test of the convergence of the full 
correction procedure.  
Normally I use $L=40 \hmpc$ and $L=10 \hmpc$ boxes with $N=256^3$, and
assess the effect of limited box size and resolution by comparison with
$L=10 \hmpc$, $N=64^3$ (the result of this standard method is shown as 
red lines highlighted by triangles), but in this figure I also show the 
results of 
applying the same correction procedure using an $L=80 \hmpc$ and 
an $L=20\hmpc$ 
box, both with $N=512^3$, and an $L=20\hmpc$ box with $N=128^3$
(black lines and squares).  This tests the assumption that the resolution 
correction can be extrapolated to large scales, even though it is estimated 
using
an insufficiently large $10 \hmpc$ box, and that the rather large box size 
correction that is needed ($\sim 30$\% along the line of sight) is reasonably
accurate.  The results are generally good, with agreement to almost 10\% in
the power for $0.5~\ihmpc < k < 10~\ihmpc$ [for $k<0.5~\ihmpc$, statistical
errors obscure the comparison].

\subsection{\lya\ Forest Bias}

The preceding calculations give a method for computing the \lyaf\ flux
power on scales smaller than $\sim 40 \hmpc$ (although many simulations
would be required in order to reduce the statistical error for the 
lowest $k$ modes).  This is sufficient for past one-dimensional power
spectrum measurements, which
probed scales less than $\sim 20 \hmpc$ 
(Croft \etal 1999, 2000; McDonald \etal 2000), but upcoming SDSS 
one-dimensional results
will reach $\sim 80 \hmpc$ (limited by uncertainties in the quasar 
continuum; L. Hui, private communication), and measurements using 
multiple lines of sight may go 
even farther.  
In order to extend the calculations of the \lyaf\ power to larger
scales than we can conveniently reach directly through simulations,
it is useful to revert to the ``linear theory with bias'' picture that
is usually used to describe large-scale clustering of galaxies
(e.g., Blanton et al. 2000; Benson et al. 2000; Cen \& Ostriker 2000; 
Somerville et al. 2001).
Within this picture we may also gain a deeper understanding of the 
$\vk$ dependence of the flux power spectrum that we see in Figure 
\ref{fullprocconv}.

We seek to determine the relationship between fluctuations in the 
mass density
field smoothed in three-dimensions on some large scale, $\sdelta$, and
fluctuations in the \lyaf\ transmitted flux field similarly smoothed,
$\sF$.  Furthermore, in contrast to the case of galaxies, the non-linear
transformation applied to the \lyaf\ field after the transformation into 
redshift space [i.e., $\exp(-\tau)$]
means that for the \lyaf\ the relationship between the large-scale 
flux fluctuations and the gradient of peculiar
velocity along the line of sight, smoothed on the same scale, 
must be determined
separately from the $\sF-\sdelta$ relation.  
I define $\eta \equiv -H^{-1} dv_\parallel / dr_\parallel$, where 
$v_\parallel$ is the peculiar velocity along the line of sight, 
and $r_\parallel$ is the distance along the line of sight, 
so $\seta$ is the smoothed gradient of the line of sight velocity.
Two bias 
parameters are needed:  one, which I will call $b_\delta$, that relates 
large-scale mass density fluctuations 
to transmitted flux fluctuations, and one, $b_\eta$, that relates 
fluctuations in the
gradient of the peculiar velocity field along the line of sight to 
flux fluctuations. 

The general formalism for bias calculations is discussed extensively 
in Dekel \& Lahav (1998),
and Appendix C of McDonald \etal (2000) discussed its application to 
the \lyaf.
The linear theory, redshift-space flux power spectrum, $P_{F,L}(k,\mu)$,
is related to the real-space mass power, $P_L(k)$, by the
familiar (Kaiser 1987) formula: 
\begin{equation}
P_{F,L}(k,\mu)= b_\delta^2~(1+\beta \mu^2)^2 P_L(k) ~,
\label{linPred}
\end{equation}
where $\beta = \Omega^{0.6}(z)~b_\eta / b_\delta$.  For $\Omega_m=0.4$ and
$\Omega_\Lambda=0.6$, $\Omega^{0.6}(z=2.25)=0.975$, so the result is very
insensitive to cosmology.
As discussed in McDonald \etal (2000), the two bias parameters come
essentially from a Taylor series expansion of $\sF$, i.e., 
$\sF(\sdelta,\seta)/\sF(\sdelta=\seta=0) \simeq 1 + b_\delta~\sdelta + 
b_\eta~\seta$, with
\begin{equation}
b_\delta =  \left. \frac{1}{\sF}\frac{d\sF}{d\sdelta} \right|_{\seta=0}~, 
\label{bdeq}
\end{equation}
and 
\begin{equation}
b_\eta = \left. \frac{1}{\sF}\frac{d\sF}{d\seta} \right|_{\sdelta=0} ~.
\label{beeq}
\end{equation}
It is these derivatives that we must measure from our simulations.

An obvious way to proceed would be to define $\sF$, $\sdelta$, 
and $\seta$ by, for example, applying a Gaussian or top-hat smoothing 
kernel to the same simulation boxes discussed in \S 3.3.  However, if I 
took this approach the simulation boxes would have to be even larger than 
the size required for convergence of the small scale structure (see 
Figure \ref{boxsize}), because they would have to be large enough to contain 
multiple internal volumes that were separately large enough to be in 
the linear regime.  Given that the simultaneous requirements on box size
and resolution already overburden my computing capabilities, I use a 
novel technique to compute the derivatives.  I take advantage of 
the fact that small-scale structure in a sufficiently large, spherical, 
over-dense or 
under-dense region evolves identically to structure in a whole universe
with the same density.  I can therefore use my numerical simulations to
simulate the evolution of structure in a large-scale perturbation by 
simply modifying the cosmological parameters that are input so they
represent the perturbation instead of the background universe.  
Using this trick, I can compute $\sF$
from $\sdelta$ and $\seta$ using the largest averaging volume that I 
can possibly simulate:  one full simulation cube.
Using multiple cubes with different values of the perturbation parameters,
$\sdelta$ and $\seta$ (I use $\sdelta=\pm 0.1$ and $\seta=\pm 0.1$),
I can compute the derivatives in equations (\ref{bdeq} and \ref{beeq})
numerically.
Note that, if I have truly reached the scale where linear theory holds, 
the size of the smoothing volume should not matter, 
since $b_\delta$ and 
$b_\eta$ are in theory independent of scale.  

I am going to assume that the growth of small-scale structure 
in a region is determined entirely by the large-scale density 
(as opposed to the complete large-scale deformation tensor).  This is
not strictly true, but it greatly simplifies the calculations.  
In order to check this assumption, it is necessary
to run simulations with anisotropic expansion.

Varying $\sdelta$ for a simulation cube still requires re-running the 
simulation,
to correctly account for the dynamical effect of changes in the mean density 
on the growth of small scale structure.  In practice this means that,
using the same set of initial particle positions and velocities, 
I evolve the simulation forward to expansion factor 
$a' = a (1-1/3~\sdelta)$, where $a$ is the original expansion factor, 
and use matter density parameter $\Omega_m' = (1+5/3~\sdelta)$, where 
originally $\Omega_m \simeq 1$ (I correctly 
account for the presence of the cosmological constant and
deviation of $\Omega_m$ from 1, but this is irrelevant in practice 
at $z>2$).  $\sF$ is simply the mean flux averaged over the
entire modified box.  The value of $\sF$ changes relative to the original
simulation, not only because of the change in overall
mean baryon density, but also because of the change in the small-scale 
distributions of densities and peculiar velocities. 

Since I am ignoring the possibility that the anisotropic expansion of 
a large-scale region is important to the growth of small-scale structure
within it,
varying $\seta$ (at fixed $\sdelta$)
within a simulation cube, for the purpose of constructing \lyaf\ spectra, 
is basically trivial. 
All optical depths are divided by a factor 
$(1-\seta)$, and the width of the thermal 
broadening kernel (i.e., $T^{1/2}$), measured in mesh cells, 
is divided by the same factor (bulk velocities measured in mesh cells
are unchanged because the overall density parameter, i.e., $\Omega$
within the perturbation, is unchanged).
This procedure is equivalent to using $(1-\seta) H(z)$ in place of $H(z)$
when creating spectra. 

Applying these transformations to a set of eight $L=40 \hmpc$, $N=256^3$ 
simulations,
I find $b_\delta=-0.1511\pm 0.0006$ and $b_\eta=-0.1722\pm0.0005$,
or, in terms of parameters whose relevance to the AP test is clearer, 
$b_\delta^2=0.0228\pm0.0002$, and $\beta=1.112\pm0.005$.
Good statistical precision can be achieved using relatively few simulations
because the parameters are being computed as differences between simulations
with identical initial conditions.  To check the convergence with box
size, I run $L=20 \hmpc$, $N=128^3$ simulations, finding 
$b_\delta^2=0.0259\pm0.0005$, and $\beta=1.052\pm0.013$.
The 6\% error in $\beta$ (equivalent to 7\% error 
in the ratio of
power along to across the line of sight) is acceptable for the present
purpose.  The 14\% error in $b_\delta^2$,
which leads to an isotropic error on $P_F(\vk)$,
would be disturbing if my purpose was to measure the amplitude of the 
primordial density fluctuations, but is not a problem for the AP test.
Finally, I test the convergence using an
$L=80 \hmpc$, $N=512^3$ simulation, finding agreement with $L=40 \hmpc$
to better than 1\% for both $b_\delta^2$ and $\beta$.

I correct the bias parameters for the limited resolution of 
the $L=40\hmpc$
simulations by applying the same corrections that are
used for the power spectrum in the lowest $k$ bins, i.e.,
by solving the following two equations for ${b'}_\delta^2$ and
$\beta'$: 
\begin{equation}
{b'}_\delta^2 = b_\delta^2 \frac{P_{F,10,256}(k_{{\rm min},10}, \mu=0)}
{P_{F,10,64}(k_{{\rm min},10}, \mu=0)}~,
\end{equation}
and 
\begin{equation}
{b'}_\delta^2 (1+\beta')^2= b_\delta^2 (1+\beta)^2 
\frac{P_{F,10,256}(k_{{\rm min},10}, \mu=1)}
{P_{F,10,64}(k_{{\rm min},10}, \mu=1)}~.
\end{equation}
The result is ${b'}_\delta^2=0.0173 \pm 0.0003$ 
and $\beta'=1.580\pm 0.022$. 
I check these bias correction factors by recomputing them using $L=20\hmpc$, 
$N=512^3$
box and a similar sized $128^3$ box, and find that the correction to $\beta$
changes by only 1\%, while the correction to $b_\delta^2$ changes by only 4\%,
indicating that my results on the largest scales have converged.

In conclusion, I have achieved something remarkable:  a direct calculation
of the linear theory bias parameters of the \lyaf, which can be used to 
extend the results of numerical simulations for comparison with data on 
arbitrarily large scales (although the importance of anisotropic 
large-scale expansion remains to be investigated).  
I demonstrated that my calculations have 
truly reached the linear regime, i.e., the bias parameters don't change
with increasing scale.  The results of this subsection and \S 3.3 suggest
that combinations of $L=40 \hmpc$, $N=512^3$ simulations with smaller 
box size simulations ($L\sim 20\hmpc$) to compute resolution corrections 
should be sufficient
to compute the \lyaf\ power spectrum for all $\vk$, to much better than 
10\% (as demonstrated in Figure \ref{fullprocconv}, my method is 
limited to $\sim 10$\% accuracy using $256^3$ 
simulations because the
largest fully resolved simulation I can run is only $L\simeq 10\hmpc$).  
With the addition of
fully hydrodynamic simulations to compute corrections to the HPM 
approximation, we should be fully prepared to interpret the clustering 
in large future data sets.  

\subsection{Analytic Formula for $P_F(\vk)$}

Equation \ref{linPred} provides a convenient analytic description of 
the power spectrum at very small $k$, but it is useful to have an
analytic formula that can describe the power for all $\vk$.
Since no such formula has been successfully derived even in the simpler 
case of the dark matter power spectrum, I follow the usual strategy of
fitting a parameterized formula to the simulation results.
The three-dimensional real-space power spectrum of the dark matter, 
in CDM models, and on the 
scale of the \lyaf, 
increases above the linear prediction with increasing $k$ 
(e.g., Ma 1998), so we might guess that the \lyaf\ flux power will behave
similarly, although there is no guarantee that it will.
Non-linear peculiar velocities should suppress the power along the line of
sight [the ``fingers of god'' effect, see Jing \& Boerner (2000) for 
the state of the art].  In the \lyaf, the addition of pressure
and thermal broadening will further modify the high-$k$ power.

Fortunately, my simulations show that the complicated transformation 
to \lyaf\ transmitted flux preserves qualitatively the
features expected for dark matter, so it is easy to guess a working fitting
formula.  I use the following general form:
\begin{equation}
P_F(k,\mu)= b_\delta^2 (1+\beta \mu^2)^2 P_L(k)~ D(k,\mu)~,
\label{fitform}
\end{equation}
where 
\begin{equation}
D(k,\mu) \equiv 
\exp\left\{\left[\frac{k}{k_{NL}}\right]^{\alpha_{NL}} - 
\left[\frac{k}{k_{P}}\right]^{\alpha_P} - 
\left[\frac{k_\parallel}{k_V\left(k\right)}\right]^{\alpha_V} \right\}~, 
\label{Deq}
\end{equation}
and $k_V(k)=k_{V0}~(1+k/k'_V)^{\alpha'_V}$.
The first term in the exponential allows for the isotropic 
increase in power due to 
non-linear growth, the second term for the isotropic suppression by 
pressure, and the third for the suppression by non-linear peculiar velocities
and temperature along the line of sight.
[The dependence of $k_V$ on $k$ is motivated by the finding of
Jing \& Boerner (2000) that the smoothing kernel associated with non-linear
peculiar velocities cannot be written as a simple function of 
$k_\parallel = \mu k$ alone.  I also find this result for the \lyaf.]

\begin{figure}
\plotone{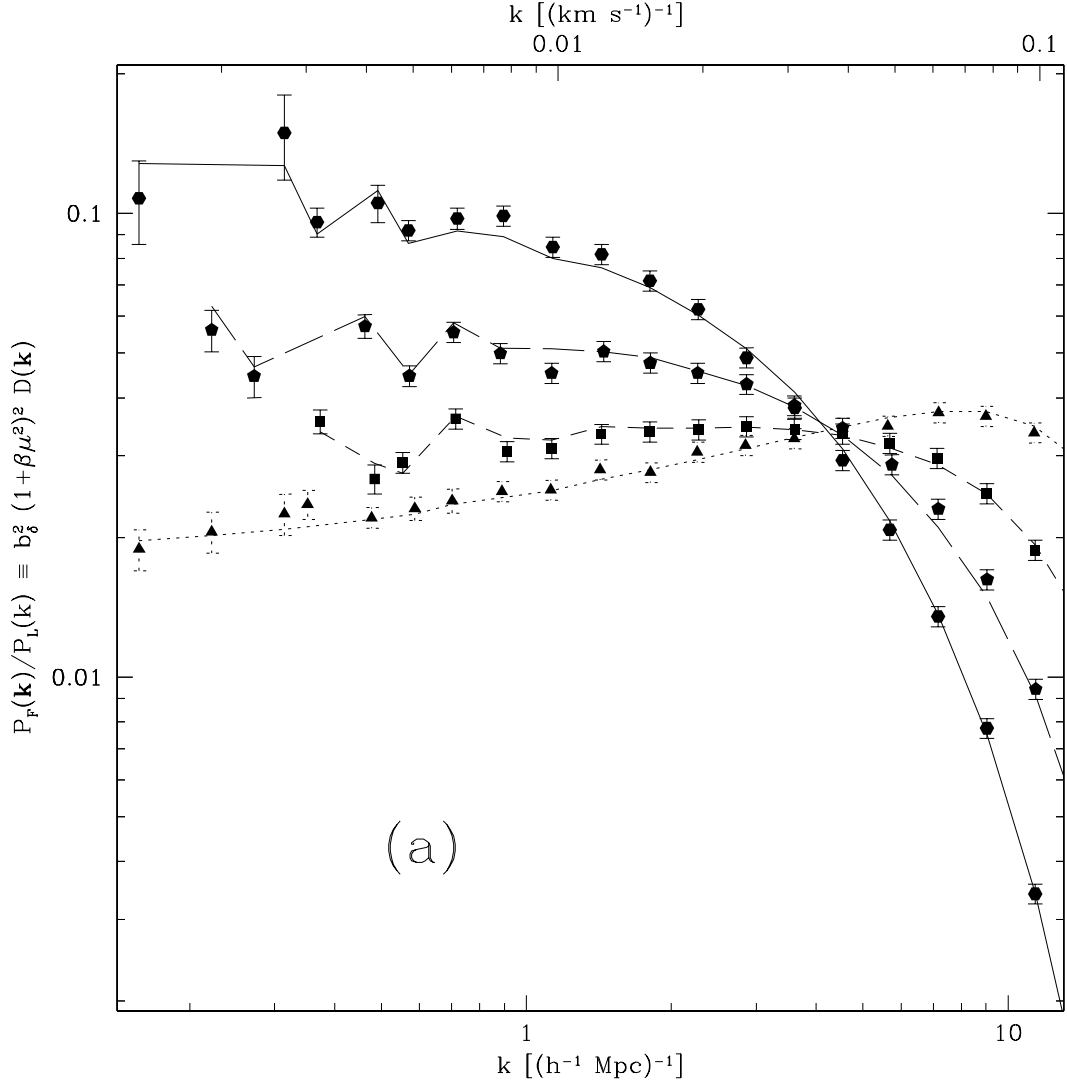}
\caption{Results for the flux power spectrum.
In (a), the {\it triangles, squares, pentagons,
and hexagons} show, for $\mu$ bins 0-0.25, 0.25-0.5, 0.5-0.75, and 0.75-1.0, 
the ratio of the \lyaf\
power to the linear theory (real space) mass power at the the same $\vk$.  
The lines show my analytic
formula, using the computed values
for $b_\delta^2$ and $\beta$, with $D(\vk)$ fit to the simulation points.
(b) is similar except $b_\delta^2 (1+\beta \mu^2)^2$ has been divided out,
leaving only $D(k, \mu)$.
}
\label{centfit}
\end{figure}

As an example, I fit equation (\ref{fitform}) to the results of my
standard $P_F(\vk)$ computation, using the values of $b_\delta^2=0.0173$ and
$\beta=1.58$ obtained by the procedure described in \S 3.4.
The error bars I use for the fit are obtained by computing the dispersion 
between
the power spectrum results from eight simulations with different random 
initial conditions.
To prevent the highest $k$ points from completely dominating the fit, I 
have set 
a minimum size for the error bars of 5\%.  
The fit is very good, better than 5\% or the statistical 
errors, in the sense that $\chi^2/\nu=0.8$.   
The fitted parameters are:  
$k_{NL}=6.77~\ihmpc$,
$\alpha_{NL}= 0.550$,
$k_P=15.9~\ihmpc$,
$\alpha_P= 2.12$,
$k_{V0}=0.819~\ihmpc$,
$\alpha_V= 1.50$,
$k'_V= 0.917~\ihmpc$, and 
$\alpha'_V= 0.528$.
Figure \ref{centfit}(a) shows the results of the fit 
with the initial real space
power spectrum divided out, i.e., $P_F(\vk)/P_L(k)$,
and Figure \ref{centfit}(b) shows just the small-scale kernel,
$D(k,\mu)=P_F(\vk)/\left[b^2_\delta \left(1+\beta \mu^2\right)^2
P_L\left(k\right)\right]$.
\begin{figure}
\plotone{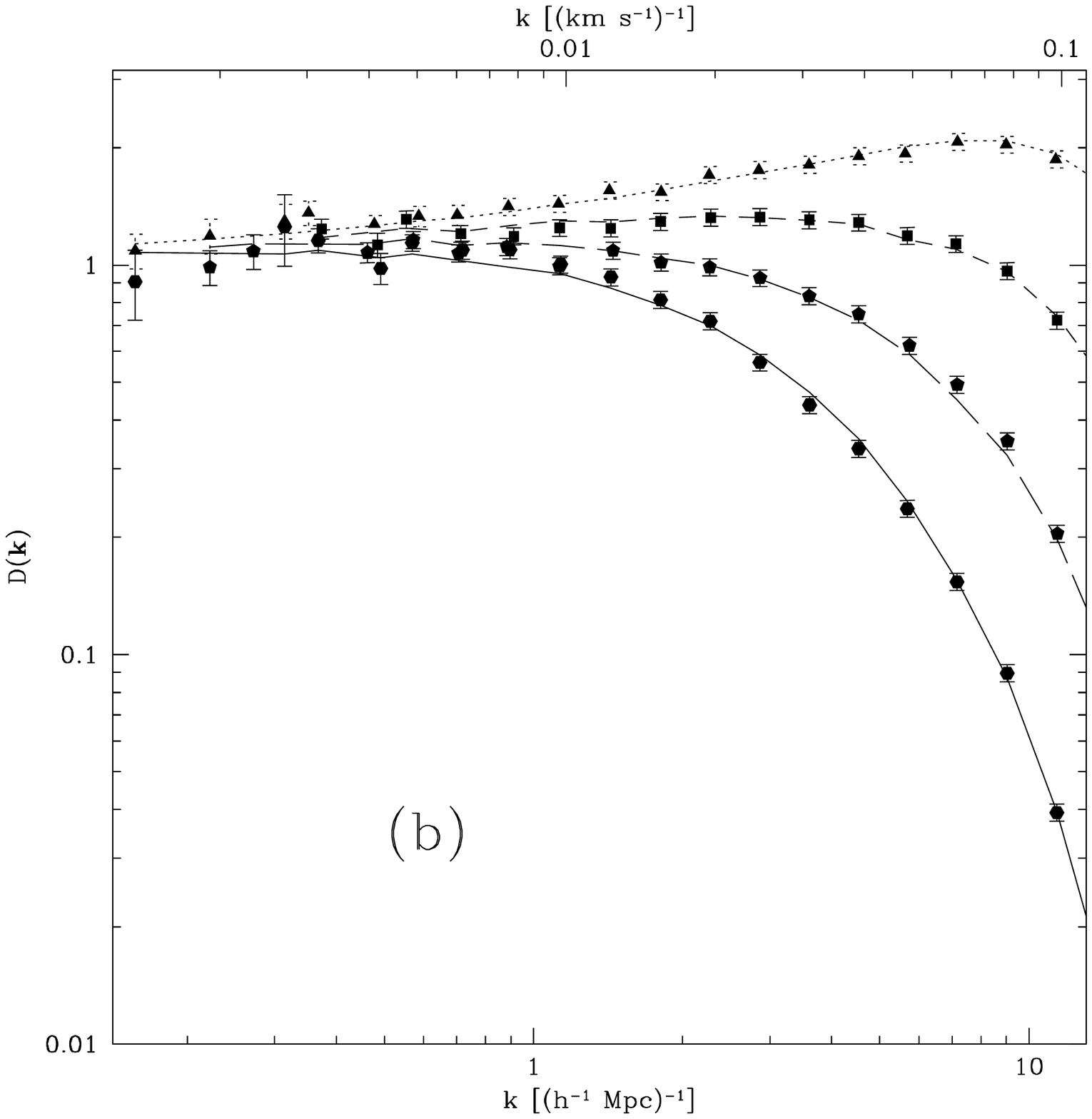}
\end{figure}
These figures summarize the primary result of this section --- 
I have established
a procedure for computing the \lyaf\ power spectrum for all $\vk$ from
HPM simulations, with reasonably well understood, if not completely
eliminated, numerical errors.

In the next section I will simply repeat the procedures described in 
this section for different
values of the cosmological and \lyaf\ model parameters.  I will use
the fitting formula presented here as a convenient way to present the
results of many different models without giving large tables of binned 
power spectrum results. In all cases I find good agreement ($\chi^2/\nu<1$)
between the power spectrum points and the best fit.
Note that I have no way to compute the parameters of the fitting formula 
from the model parameters, other than running the necessary simulations
and doing the fit, so the formula is only useful for interpolating 
between models,
not for extrapolating to models outside the basic parameter space in which
I provide results.  Equation (\ref{fitform}) also eliminates the problem of
interpolating between many power spectrum points, binned in $k$ and $\mu$,
when computing quantities which are integrals over the three-dimensional 
power spectrum, such as the correlation function or the one-dimensional
power spectrum.

\section{PARAMETER DEPENDENCE OF THE POWER SPECTRUM}

In order to carry out a complete analysis of \lyaf\ data, it will
be necessary to perform the kind of multi-parameter, maximum-likelihood 
analysis that is commonly applied to the CMB power
spectrum (e.g., Tegmark \& Zaldarriaga 2000), involving a many-dimensional
grid of \lyaf\ model predictions (Zaldarriaga \etal 2000).  Here I am
content to compute the derivatives of $P_F(\vk)$ with respect
to the parameters at a typical point in parameter-space.
These derivatives
show the basic trends, and allow me to perform Fisher matrix
calculations of the expected parameter constraining power of
future observations (see, for example, Eisenstein, Hu, \& Tegmark 1999).

I choose to study a model in which the flux power spectrum depends on 
five free parameters:
$A_1$, the value of $\Delta^2_L(k) \equiv k^3 P_L(k)/(2 \pi^2)$
at $k_1=2 \pi \ihmpc$,
$n_1$, the power law index of the power spectrum 
at $k_1$, $T_{1.4}$, $\gmo$, and $\bF$, which 
together form the parameter
vector ${\mathbf p}=(\bF,~T_{1.4},~\gmo,~A_1,~n_1)$.  
I use a CDM shape for the power spectrum, not a true power law, but 
for simplicity I do not allow the shape to change, other than by an overall
tilt controlled by $n_1$.  My results can be applied to any cosmological
model in which the power spectrum can be approximated by the CDM shape, 
by computing $A_1$ and $n_1$ for the power spectrum of the model.

\begin{figure}
\plotone{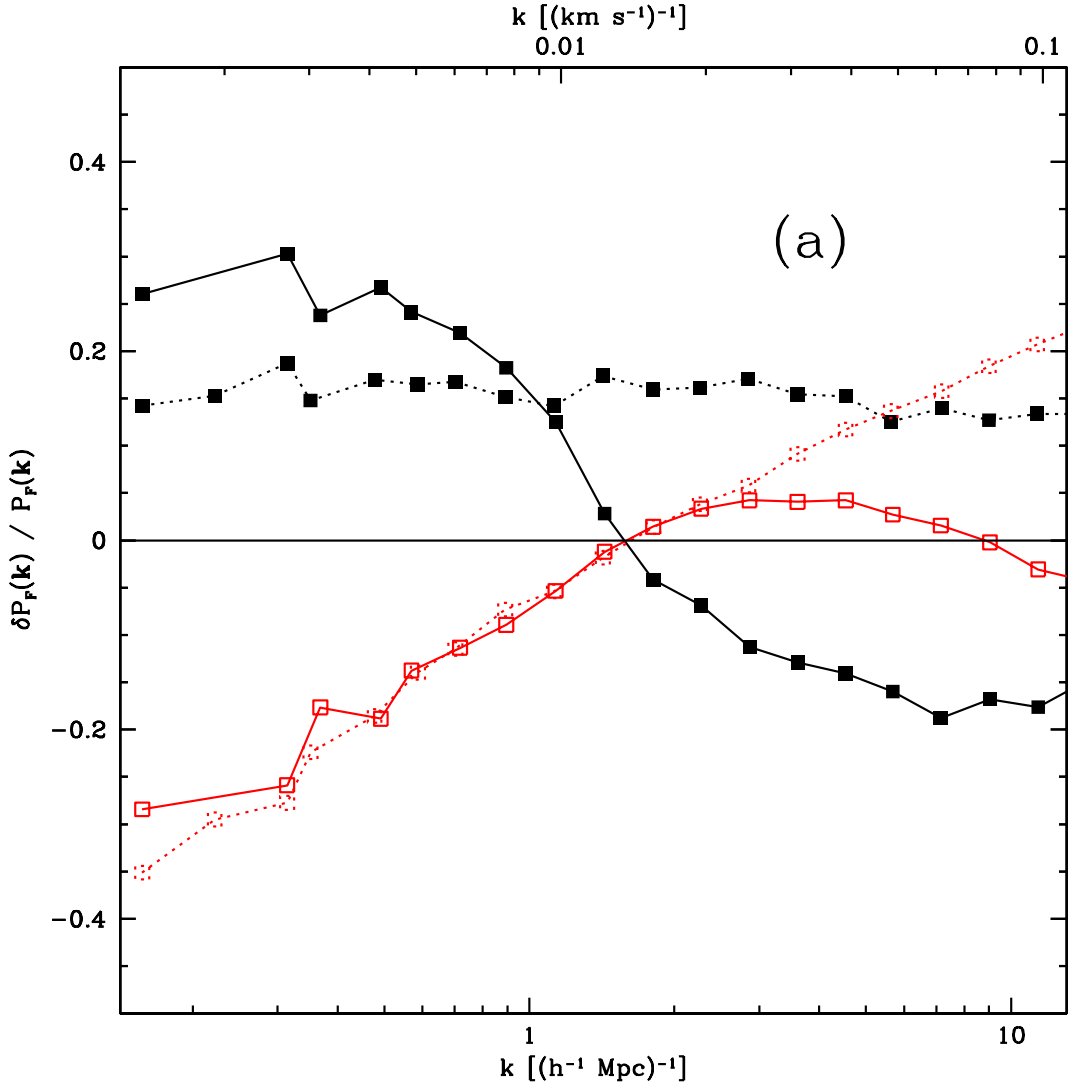}
\caption{Fractional variation of $P_F(\vk)$ with variation of the 
model parameters.  
The {\it solid}
lines are power along the line of sight, the {\it dotted} are transverse
power.  
As discussed in the text, $\delta P_F$ is the difference between results
for a positive and negative variation of each parameter, i.e., 
$\delta P_F \equiv P_F(p+\delta p)-P_F(p-\delta p)$, 
where $p$ is the parameter.
In (a), {\it black lines} highlighted by {\it solid squares} show  
variation of $A_1$, by $\pm 0.29$, 
and {\it red lines} ({\it open squares}) show  
variation of $n_1$, by $\pm 0.1$.
In (b), {\it black lines} ({\it solid squares}) show  
variation of $T_{1.4}$ by $\pm 2000$ K, 
and {\it red lines} ({\it open squares}) show  
variation of $\gmo$ by $\pm 0.1$.
In (c), {\it black lines} ({\it solid squares}) show  
variation of $\bF$ by $\pm 0.025$, 
and {\it red lines} ({\it open squares}) show  
variation of $z$ by $\pm 0.25$ (see text).
}
\label{derivs}
\end{figure}

I vary each of the parameters and
compute the flux power spectrum as described in \S 3.  The central
set of parameters is 
${\mathbf p}_0 = \left(0.8,~20000,~0.5,~1.38,~-2.58\right)$
(these values for $A_1$ and $n_1$ are equivalent 
to $\sigma_8=0.79$ and $n=0.95$
for the flat, $\Lambda$CDM model, with $\Omega_m=0.4$ and $h=0.65$).
In Figures \ref{derivs}(a,b,c) I plot the quantity
\begin{equation}
\frac{\delta P_F}{P_F}\left(\delta {\mathbf p}\right) \equiv
\frac{P_F\left({\mathbf p}_0+\delta {\mathbf p}\right)-
      P_F\left({\mathbf p}_0-\delta {\mathbf p}\right)}
      {P_F\left({\mathbf p}_0\right)}~,
\end{equation}
where $\delta {\mathbf p}$ is some variation of the parameters. 
The values of $\delta {\mathbf p}$ I choose are intended to represent
roughly the current level of uncertainty in each parameter.

When we look at $\delta P_F/P_F$ for different parameter variations, 
we will be particularly interested in any parameter variations that 
could lead to substantial variations in the ratio of the power along
the line of sight to the power across the line of sight (i.e., variations
in $\beta$ for small $k$), because these could lead to degeneracy 
between the parameter
being varied and the transverse scale factor $f(z)$ that is measured in the
AP test.  Fortunately, the one-dimensional power spectrum can be measured
very accurately from single lines of sight and used to constrain the
model parameters, so we only need to worry if there is model dependence 
of the power spectrum anisotropy for relatively small variations in the
parameters.  

In Figure \ref{derivs}(a), the black lines highlighted by 
filled squares show the variation in power spectrum amplitude,
$\delta A_1 / A_1 = 0.29/1.38 = 0.21$, with all of the other
parameters fixed.
\begin{figure}
\plotone{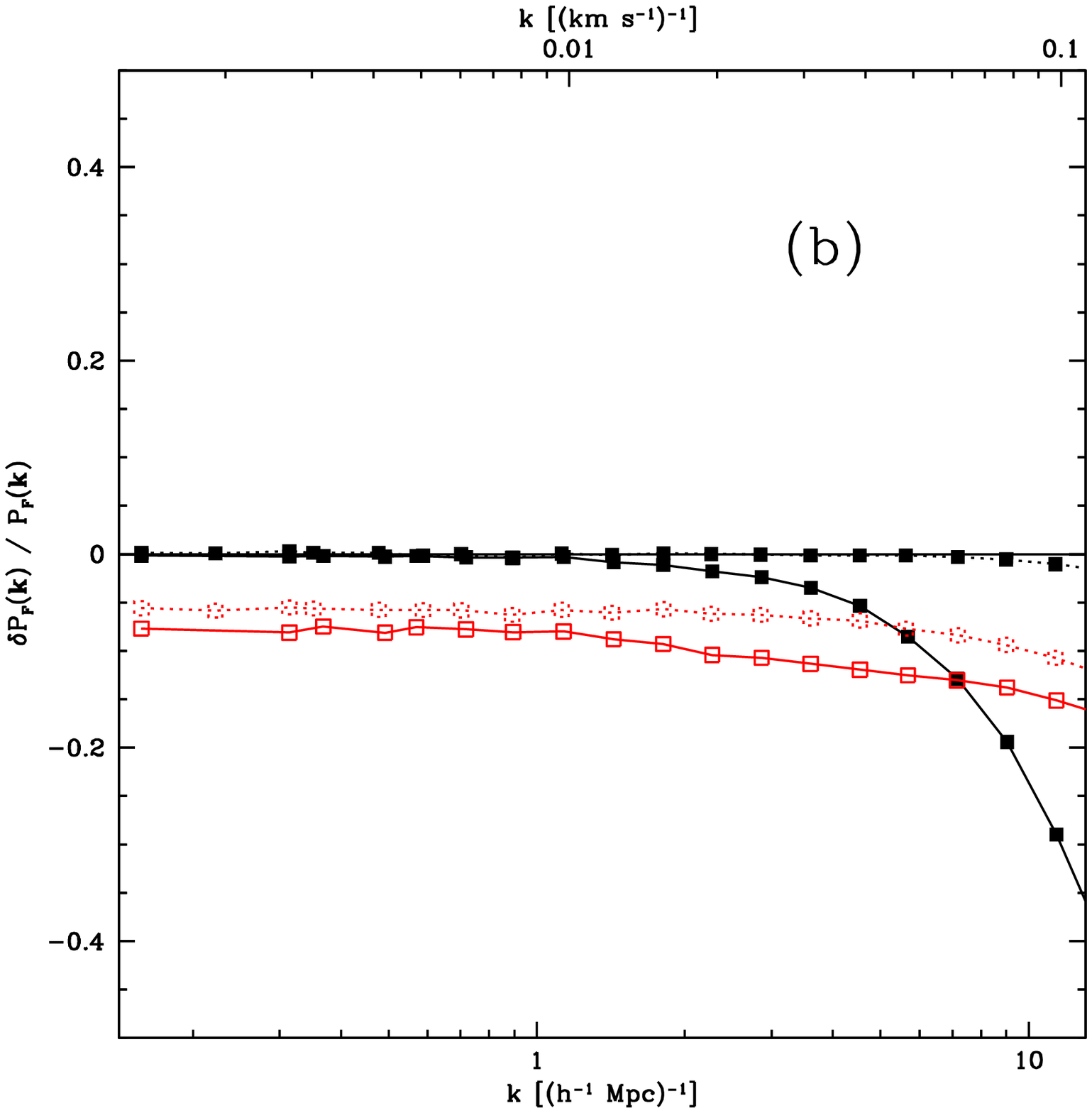}
\end{figure}
\begin{figure}
\plotone{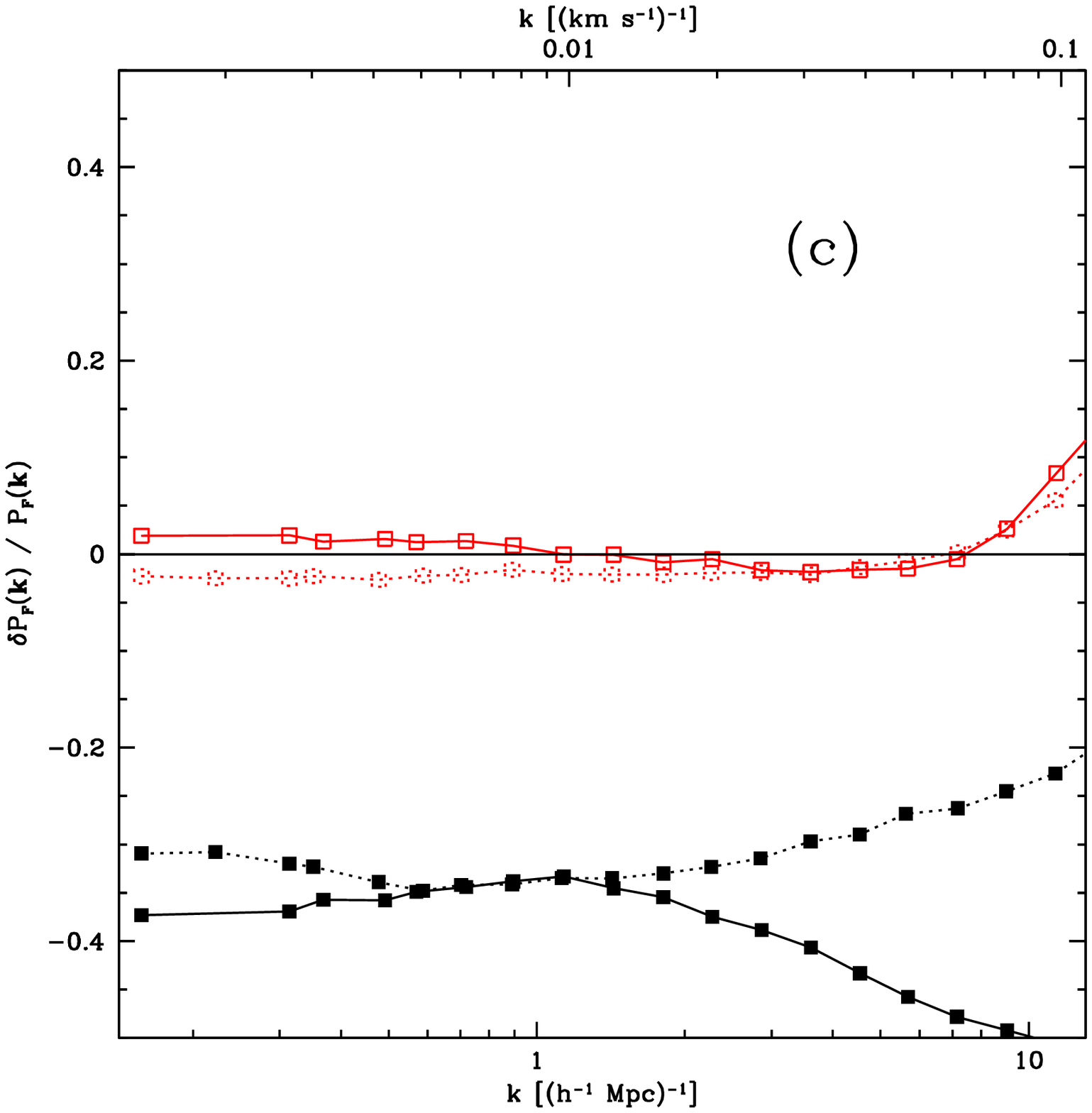}
\end{figure}
We see the expected increase in large-scale power both along and across
the line of sight, with the increase in flux power significantly less than 
proportional to the increase in initial mass power, as found by 
Croft \etal (1999; 2000), and McDonald \etal (2000).  For 
$k\gtrsim 1-2 \ihmpc$, it is interesting to note that increasing the mass
power actually decreases the flux power along the line of sight, presumably
by increasing the power suppression by non-linear peculiar velocities.   
The variation of $\sim \pm 5$\% in the ratio of power along to across the 
line of sight on
large scales, equivalent to a similar variation 
in $\beta$, is actually not a problem for the AP test
because the measurements by McDonald \etal (2000) and
Croft \etal (2000) already constrain the power spectrum amplitude
to better than $\sim 20$\%.  The variation in power spectrum anisotropy will 
become more of a problem for quasar pairs with separations less 
than a few arcminutes, where the power suppression by non-linear peculiar
velocities is important.

The red lines, highlighted by open squares, in Figure \ref{derivs}(a) show the 
variation $\delta n_1=0.1$.  (Due to imperfect planning, this variation
is at fixed $\sigma_1$, not fixed $A_1$, where $\sigma_1$ is the rms
fluctuation of the linear theory mass field in spheres with radius $1\hmpc$;
however, the difference is not important to the qualitative presentation in 
the figure.) At low $k$, the effect of changing $n_1$ appears to be a 
straightforward, isotropic, change in the slope of the flux power spectrum.
At high $k$, where increasing $n_1$ means increased initial power, $P_F(\vk)$
is again suppressed along the line of sight.

Figure \ref{derivs}(b) shows variations of the temperature-density
relation, $\delta T_{1.4}=2000$ K (black lines with solid
squares) and $\delta(\gmo)=0.1$ (red lines with open squares) 
(actually, $\gmo$ was only varied by $\pm 0.05$ in the simulations
so $\delta P$ has been multiplied by 2).
The variation with $T_{1.4}$ is just what might be expected from the
corresponding change
in thermal broadening.  The thermal broadening   
suppresses the optical depth power by $\exp(-k^2 T)$,
so, for $2~\delta T= 4000$ K, the increased suppression 
is by $\exp(-k^2 ~4000 {\rm K})=\exp[-(k~5.75 \kms )^2]$, or a factor 0.72 
at $k=0.1\ikms$ --- almost precisely the suppression seen in the Figure,
despite the non-linear transformation from optical depth to transmitted
flux.  

The dependence of $P_F(\vk)$ on $\gmo$ is also simple
to understand, through equation (\ref{taueq}).
Since $\tau\propto \Delta^\beta =(1+\delta)^\beta$, 
with $\beta=2-0.7(\gmo)$, we expect that an increase in $\gmo$ will
decrease $\beta$ and thus decrease the fluctuations in optical depth
for given fluctuations in $\delta$.  A naive expansion assuming 
small $\delta$ predicts that the \lyaf\ bias is proportional to $\beta$
(Hui 1999; McDonald \& Miralda-Escud\'e 1999) which 
would lead to a decrease in power by 17\% for $2 \delta (\gmo)=0.2$.
The actual decrease in Figure \ref{derivs}(b) is less than half this 
prediction, which is reminiscent of the weaker than proportional 
dependence of the flux power on mass power.  Note that the choice of
$\Delta=1.4$ for the normalization of the temperature-density relation
appears to have the intended consequence of making the effective
smoothing by thermal broadening mostly independent of $\gmo$
(see McDonald et al. 2001).   
     
The final parameter, $\bF$, is varied by $\pm 0.025$ in 
Figure \ref{derivs}(c) 
(black line with filled squares; 
$\bF$ was actually varied by $\pm 0.05$
in the simulations, so $\delta P_F$ has been divided by 2).
The decreasing power with increasing $\bF$ is expected from the
one-dimensional results of Croft \etal (1999) and 
McDonald \etal (2000).  Fortunately
for AP test purposes, the large-scale anisotropy does not depend 
sensitively on $\bF$.  

Now that we know how $P_F(\vk)$ depends on the parameters, at 
fixed redshift, it is interesting to see if we can treat a change
in $z$ as a simple rescaling of the parameters (as assumed by
McDonald \etal 2000). 
To create the red curve (with open squares) in Figure \ref{derivs}(c),
I took outputs from the simulations at $z=2.0$ and $z=2.5$, instead
of the usual $z=2.25$.  Both were analyzed using the usual values
of $\bF$, $T_{1.4}$, and $\gmo$.  Two obvious changes with redshift 
have been scaled away:  The change in linear growth factor, approximately 
$\propto (1+z)^{-1}$, can be treated as a change in power spectrum 
amplitude, $A_1$, so I have subtracted 
\begin{equation}
\frac{d P_F}{d A_1} \frac{d A_1}{d z}~2 \delta z \simeq
\frac{\delta P_{F,A_1}}{2~\delta A_1} 
\left[\left(\frac{1+2.25}{1+2.5}\right)^2-\left(\frac{1+2.25}{1+2.0}\right)^2 
\right] A_1
\end{equation}
from the measured $\delta P_{F,z}$ (where I am using $\delta P_{F,p_i}$
to indicate the variation of $P_F$ in the simulations with parameter $p_i$).
A much less important term arising from the change in the temperature 
measured in comoving coordinates is also subtracted.  The resulting curve
shows that the parameter scaling approximation is close to perfect, 
except for a small deviation at the highest $k$.  The primary remaining known 
error in the approximation is that the specified thermal history of the
gas, which sets the pressure in the simulations, is offset by the difference
in the two redshifts.  
The increase in power at high $k$ might be explained by this
if the effective smoothing scale is smaller than the Jeans scale, 
but increasing towards it over time, as predicted by Gnedin \& Hui (1998).    

In order to present these results in a quantitative, compact form that may 
be useful to others, I have fitted equation (\ref{fitform}) to 
$P_F(\vk)$ for each of the parameter settings used in 
Figures \ref{derivs}(a,b,c).  Table \ref{fitparms} lists the values of 
$b_\delta^2$ and $\beta$ for each variation of the model parameters, 
and the 
parameters of $D(\vk)$ for each fit.  To save computer and organizational 
time, I do not recompute
$b_\delta$ and $b_\eta$ for each variation by the method in \S 3.4, because
each $b_\delta$ requires the running of two extra simulations.  Instead,
the new values are set by solving
${b'_\delta}^2=b_\delta^2~P'_F(k_{{\rm min}}, \mu=0)/
P_F(k_{{\rm min}}, \mu=0)$,
and ${b'_\delta}^2 (1+\beta')^2=b_\delta^2 (1+\beta)^2~ 
P'_F(k_{{\rm min}}, \mu=1)/P_F(k_{{\rm min}}, \mu=1)$, where the unprimed
quantities are at the central parameter values, and the primed are at the
varied parameter values (considering the convergence tests in \S 3.4, 
this shortcut should be perfectly accurate).  The agreement between the
simulation results and the fitting results is good in all cases, in the
sense that $\chi^2/\nu <1$ using error bars computed as described 
in \S 3.5.      

\section{IMPLICATIONS FOR THE AP TEST}

The primary motivation for computing $P_F(\vk)$ is to perform the 
AP test using the correlation between absorption in multiple lines
of sight.  McDonald \& Miralda-Escud\'e (1999), and Hui, Stebbins, \&
Burles (1999) discussed using the \lyaf\ AP test to 
measure $\Omega_\Lambda$, and estimated the precision that could
be obtained from hypothetical sets of data; however, these estimates
were not based on any realistic calculation of the flux correlation,
and did not address many of the relevant observational issues like the
requirements on spectral resolution and signal-to-noise ratio.  I can 
now do much better.  
Using my computed derivatives of $P_F(\vk)$ with respect to the model
parameters, I can compute the Fisher information
matrix for any hypothetical data set, and use it to find the 
smallest possible error bars on the parameters $p_i$, and the effects
of different assumptions about the data quality.

\subsection{\lya\ Forest Fisher Matrix}

A very clear discussion of the Fisher information matrix and its uses 
can be found in
Tegmark, Taylor, \& Heavens (1997, see also references therein), here
I only outline the essential points.
If we represent a data set (i.e., a set of pixels in \lyaf\ spectra) by the 
vector
${\mathbf x}$, and define the likelihood of observing ${\bf x}$ 
in a model with parameters ${\bf p}$ to
be $L({\mathbf x};{\mathbf p})$, the Fisher matrix is
\begin{equation}
F_{i j} = - \left< \frac{\partial^2 {\mathcal L}}{\partial p_i \partial p_j}
\right>~,
\end{equation}
where ${\mathcal L}=-\ln L$, and the $\left<...\right>$ brackets mean ``average
over all possible ${\mathbf x}$.''
If a maximum likelihood estimate is made of one parameter, 
$p_i$, with the others
fixed, the rms error on $p_i$ will be $1/F_{i i}^{-1/2}$.
If the other parameters are marginalized over, the error bar
on $p_i$ is $(F^{-1})_{i i}^{1/2}$.  One very useful fact about the 
Fisher matrix is that the error bars obtainable by combining multiple 
independent data sets can be estimated by simply adding up all the $F$'s.
Similarly, imposing a prior constraint, call it $\sigma_i$, on $p_i$ is a
simple matter of adding $1/\sigma_i^2$ to $F_{i i}$.

Calculating $L({\mathbf x};{\mathbf p})$ for the \lyaf\ transmitted
flux is in general 
difficult; however, for small enough $k$ we expect that the Fourier modes
will be independent and Gaussian, in which case ${\mathcal L}$ is given by
\begin{equation}
2 {\mathcal L} = \ln \det {\mathbf C} + 
\left({\mathbf x}-{\mathbf \mu}\right)^T {\mathbf C}^{-1}
\left({\mathbf x}-{\mathbf \mu}\right) + constant~,
\end{equation} 
where ${\mathbf \mu} = \left< {\mathbf x}\right>$ (not to be confused with
$k_\parallel/k$), and
${\mathbf C} = \left<\left({\mathbf x}-{\mathbf \mu}\right)
\left({\mathbf x}-{\mathbf \mu}\right)^T\right>$, i.e., 
$\mu_i$ is the mean transmitted flux at pixel $i$, and
$C_{i j}$ is the correlation between pixel $i$ and pixel $j$.
The correlation between pixels is found from my computed $P_F(\vk)$ using
equation (\ref{coreqftpow}).  
The effects of resolution and pixelization
are included by convolving $\xi_F$ with the appropriate window 
functions, and the mean squared noise level at pixel $i$ is finally added to
$C_{i i}$.   

I have tested the validity of the assumption of independent Fourier modes 
by running many simulations with 
identical parameters, but different random initial conditions, and 
comparing the
dispersion in the binned power spectrum measurements to error predictions
made by assuming independent modes.  By this test, the 
approximation appears to work well for $k \lesssim 2 \ihmpc$.
I have done a preliminary check that the AP test using pairs with 
separations 
greater than a few arcminutes is primarily sensitive to power at
$k \lesssim 2 \ihmpc$, so my Fisher matrix calculations should
be reasonably accurate.        

\subsection{Application:  SDSS Spectra}

The SDSS will obtain spectra of $\sim 100000$ quasars, making the potential
application of the \lyaf\ AP test using this data very exciting.
Fan (1999) gives predictions for the expected number
of quasars as a function of redshift and magnitude, 
which can be used to estimate the
number of close pairs that will be found in the full 10000 square degrees
of the survey.  Assuming the planned
limiting magnitude for obtaining spectra 
of $i'<19.5$\footnote{http://www.sdss.org} 
(unfortunately, as I discuss below, this limit is not being 
reached), and counting 
only the \lyaf\ region of spectra at $z\gtrsim 2.125$ (the cutoff below
which the \lyaf\ will not be observed), I estimate that SDSS would
find overlapping regions of spectra equivalent to 
\begin{equation}
N_{pair}(<\theta) \simeq 13 \left(\frac{\theta}{1'}\right)^2
\label{Npairs}
\end{equation}
pairs with equal quasar redshifts and 
complete \lyaf\ coverage, 
at separation less 
than $\theta$ (the actual number of partial pairs is of course larger
than this).  
The mean redshift of the overlapping forest is $\bar{z}\simeq 2.35$.

For simplicity, the calculation that produced equation (\ref{Npairs}) 
assumed a random distribution of quasars, 
ignoring the increase in the number of pairs because of correlation.
I estimate the effect of correlation by considering the 
number of quasars within transverse distance $R$ from a given quasar,
weighted by the fractional overlap of their \lyaf\ spectra: 
\begin{equation}
N(<R) = 2 \pi ~ \bar{n} \int_0^R r ~ dr \int_{-z_{max}}^{z_{max}} 
         \left( 1-\left|\frac{z}{z_{max}}\right|\right)
       \left[1+\xi\left(\sqrt{r^2+z^2}\right)\right] dz~,
\end{equation}
where $\bar{n}$ is the mean density of quasars, $z_{max}\sim 400\hmpc$ 
is the
maximum overlap distance along the line of sight, and $\xi$ is the
correlation function.  The fractional increase in pairs at a given
separation due to 
correlation is given by
\begin{equation}
\frac{dN/dR}{dN_0/dR}-1=\frac{
\int_0^{z_{max}} dz \left(1-z/z_{max}\right) 
\xi\left(\sqrt{R^2+z^2}\right)}
{\int_0^{z_{max}} dz \left(1-z/z_{max}\right)}~,
\label{corinc}
\end{equation}
where $N_0$ is the uncorrelated case.  
Croom et al. (2001) measure the quasar correlation function 
from 2dF data and give results in terms of a power law
$\xi(r)= \left(r / r_0\right)^{-\alpha}$.  At $\bar{z}=2.36$, they 
find $r_0 = 6.93^{+1.32}_{-1.64} \hmpc$ 
and $\alpha = 1.64^{+0.29}_{-0.27}$ (assuming a flat universe 
with $\Omega_m=0.3$).  Evaluating equation (\ref{corinc})
using the measured correlation, I find an increase of only 25\% in 
the number of pairs at separation $R=1\hmpc$ ($\theta \sim 1'$), which 
falls to 5\% at $R=10\hmpc$.  This difference is insignificant in the
following discussion.

\begin{figure}
\plotone{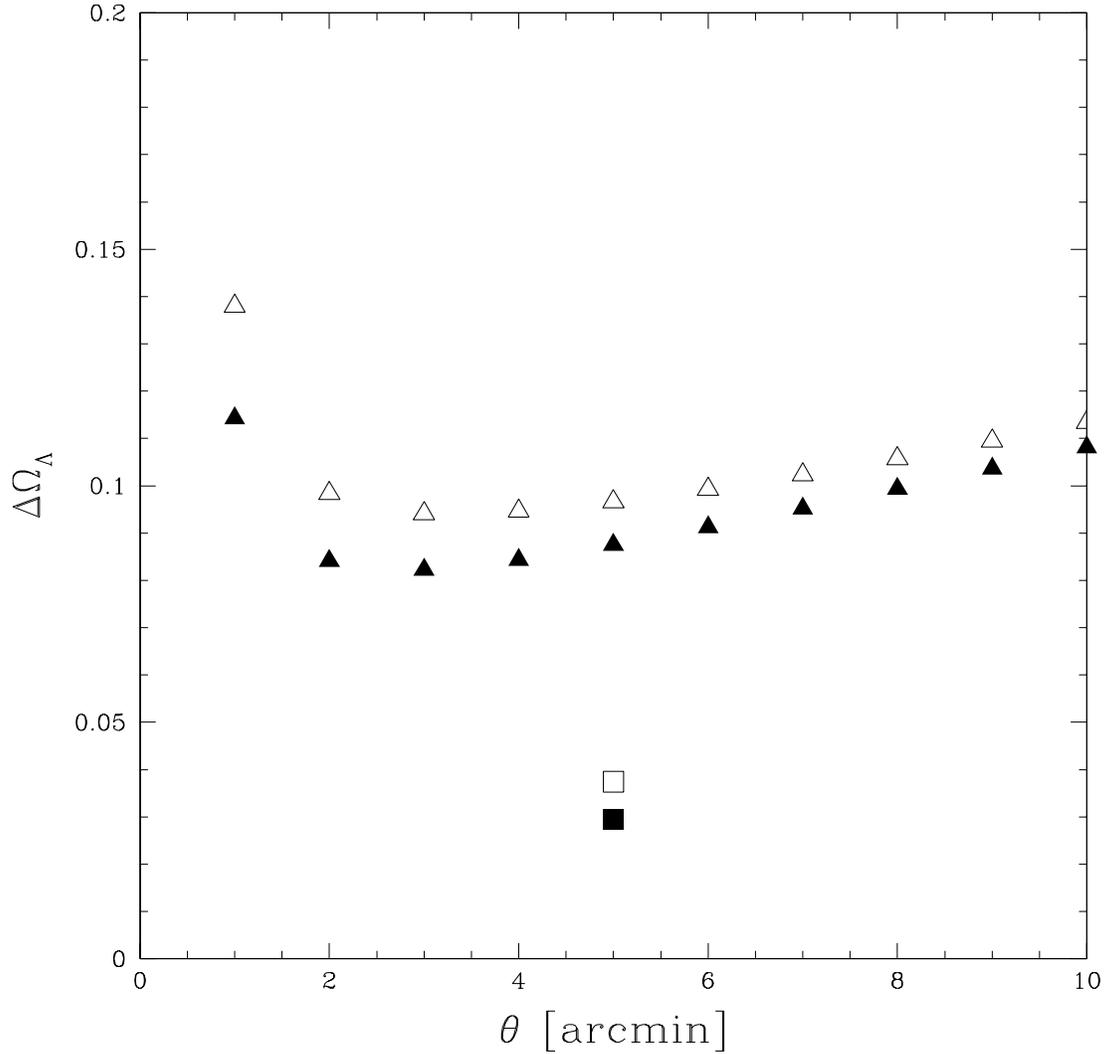}
\caption{Projected error bars on $\Omega_\Lambda$ from 
$N_{pair}(<\theta) = 13 \left(\theta / 1'\right)^2$ fully overlapping
spectra (for true value $\Omega_\Lambda \simeq 0.7$, and
assuming $\omega=-1$ and $z=2.25$).
The triangles show the constraint using only pairs in each 1 arcminute
interval of separations.  The squares show the overall error,
for $\theta < 10'$. 
The filled symbols assume that the parameters other 
than $f(z)$ are known,
the open symbols assume they are marginalized over.  In this, 
and the following figures, the horizontal positioning of the 
point showing the overall error is meaningless.
}
\label{basicsdssfcon}
\end{figure}

The SDSS signal-to-noise ratio for 1\AA\ pixels is expected to be greater
than 10 for a typical spectrum, with resolution 2000, or $\sim 2$\AA\
FWHM at $z=2.3$.  For my first Fisher matrix calculation, I will 
assume that SDSS-quality spectra can be used off the shelf, 
i.e., I assume S/N=10, 1\AA\ pixels, and resolution 2\AA.  
Figure \ref{basicsdssfcon} shows basic results for the error bars 
on $f(z)$.
(To make the results less abstract, I have translated the error on
$f(z)$, $\Delta f(z)$, into an error on the cosmological constant, 
$\Delta \Omega_\Lambda \simeq 1.25~\Delta f$, by
assuming $z=2.25$, a flat universe, $\omega=-1$, 
and $\Omega_\Lambda \simeq 0.7$.)
The squares show the estimated value of $\Delta \Omega_\Lambda$ obtained using
all of the pairs with separation less than 10' (this cutoff is arbitrary,
because the maximum separation, which will probably be set by quasar 
continuum issues, is unknown).  To compute the solid
square, I have assumed that the model parameters other than $f(z)$
($\bF$, $T_{1.4}$, $\gmo$, $A_1$, and $n_1$) are known.  This is not
unreasonable because a huge number of single spectra will be available
to constrain them.  For the open square, I have marginalized over
the other parameters assuming only the very weak constraints 
$\sigma_T=4000$ K, $\sigma_{\gamma-1}=0.3$, and $\sigma_{\bF} = 0.03$.
The triangles are the error projections
for independent measurements of $f(z)$ made after splitting the pairs into
groups by separation.  Each triangle represents the constraint using only the 
pairs in a 1' interval of separations.  The distinction between 
open and solid triangles is the same as for squares.
This figure contains the primary result of this section:  we can expect
to measure $\Omega_\Lambda$ to $\pm 0.03$ or $\pm 0.04$ using
$N_{pair}(<\theta) = 13 \left(\theta / 1'\right)^2$ fully overlapping,
SDSS-quality spectra.  As a 
consistency check, we can make measurements accurate to
about $\pm 0.1$ for each one arcminute interval of separations 
[$f(z)$ should not depend on $\theta$].  The overall
result is not very sensitive to pairs at small separations where some of my
approximations will be least accurate.

\begin{figure}
\plotone{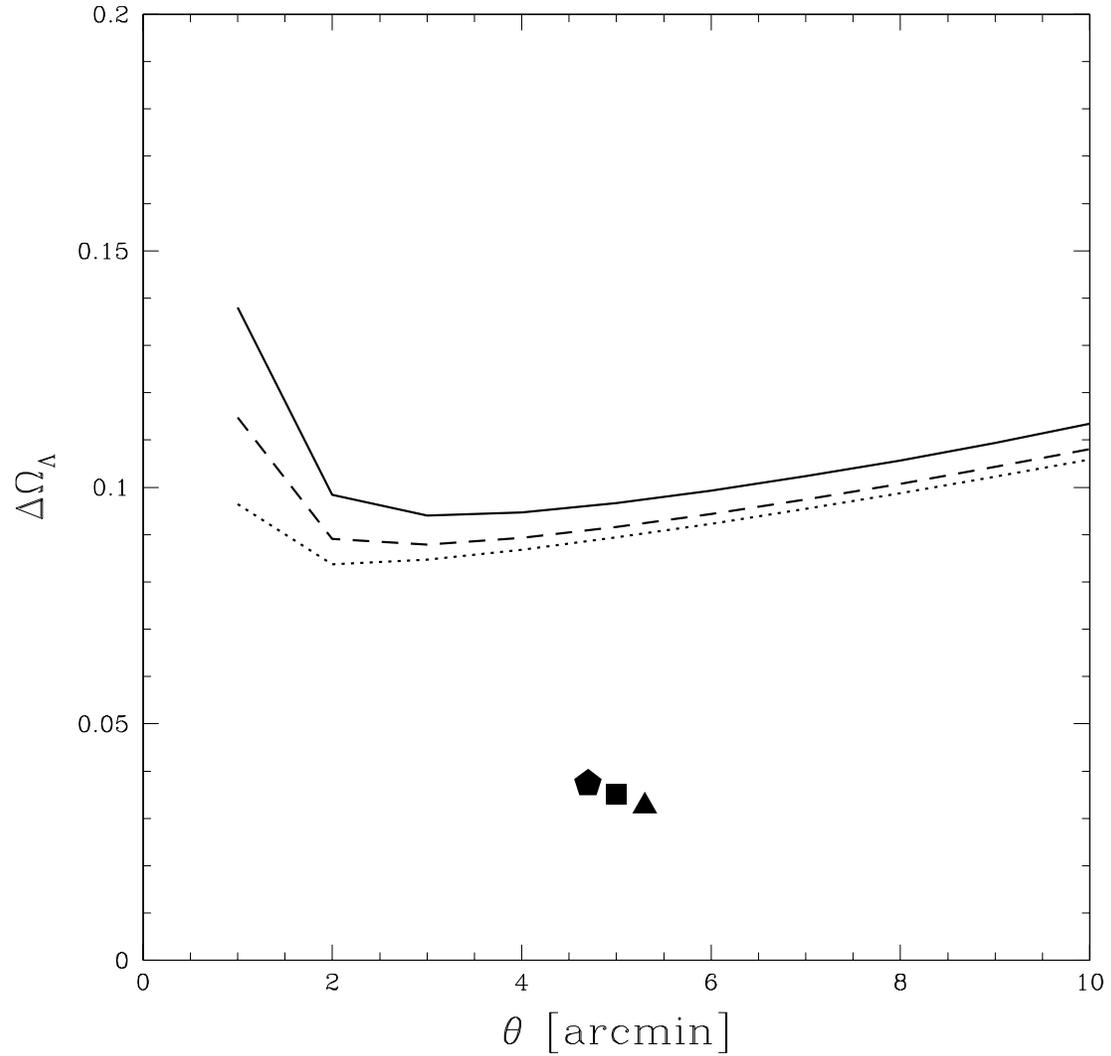}
\caption{Sensitivity of error bars to noise level.
The pentagon, square, and triangle represent S/N=10, 20, and 100, 
respectively, using all pairs with $\theta < 10'$, while the 
solid, dashed, and dotted lines show the error in 1' intervals
for the same respective S/N.
}
\label{lownoisesdssfcon}
\end{figure}

\begin{figure}
\plotone{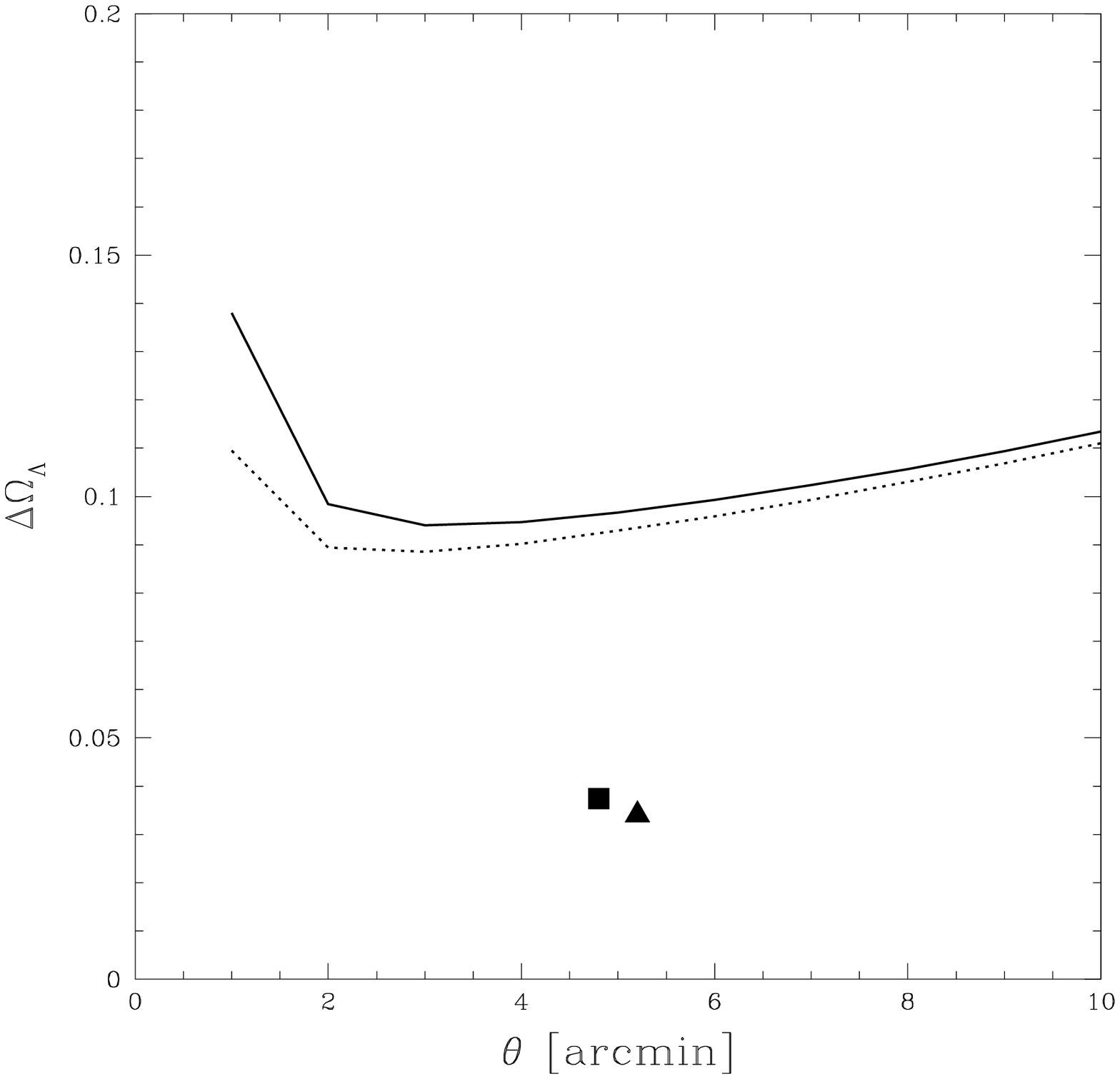}
\caption{Sensitivity of error bars to resolution.
The solid line and the square show resolution 2 \AA.  The dotted
line and triangle show 1 \AA.  Resolutions are FWHM, with pixel
sizes 1\AA\ and 0.5\AA, respectively.
}
\label{highressdssfcon}
\end{figure}

Can we improve the measurement of $f(z)$ by
taking spectra with better resolution or signal-to-noise ratio?
Figure \ref{lownoisesdssfcon} shows the error estimates with
S/N increased to 20 and 100.
Improving S/N does not improve the overall results very much,
helping most at small separations.
Figure \ref{highressdssfcon} shows the improvement with increased
resolution.
Again, the reduction in the error bars is greatest at small separations 
(not surprisingly),
but does not improve the overall result very much.  
  
\begin{figure}
\plotone{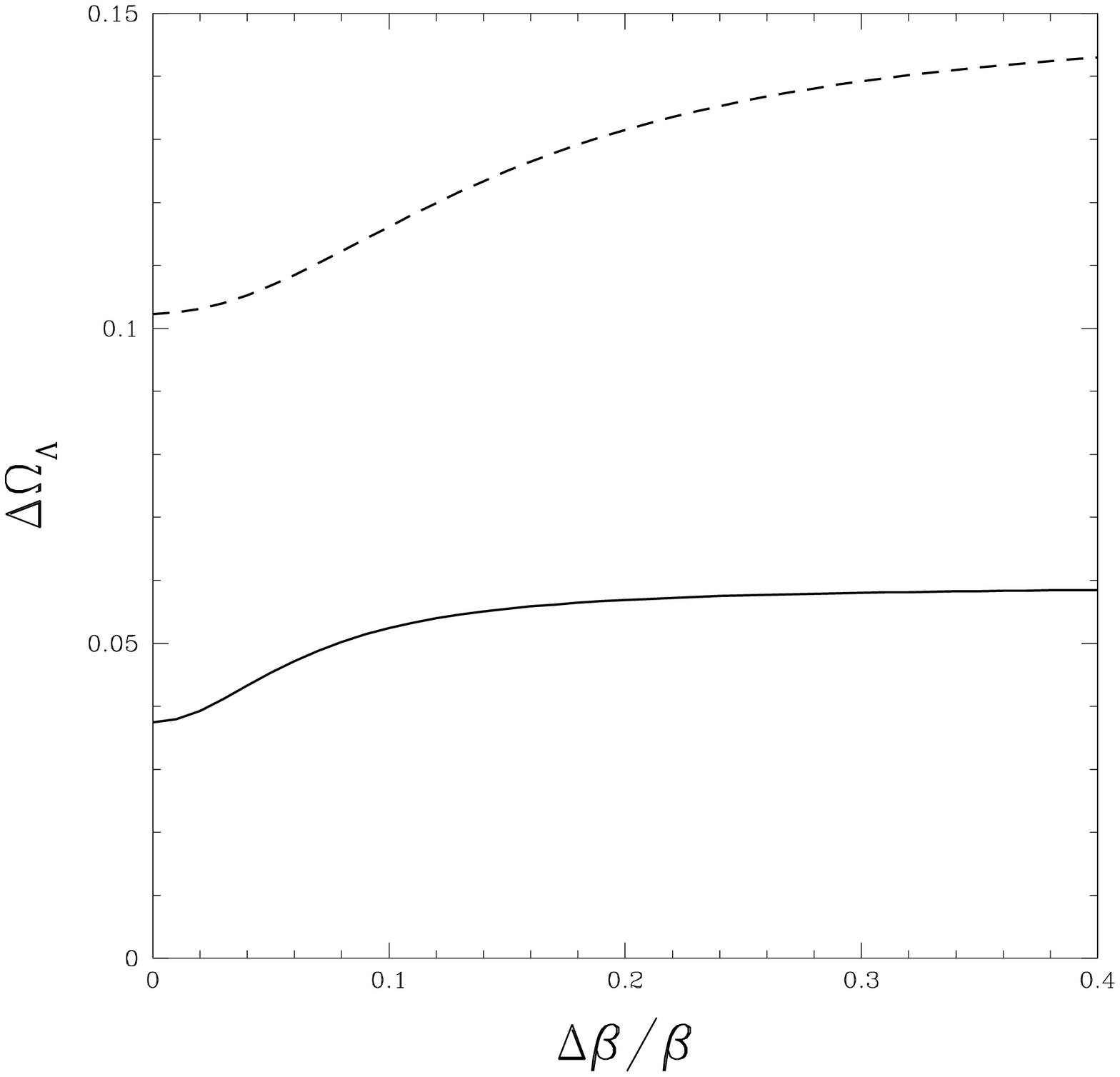}
\caption{Sensitivity of error bars to accuracy of the power spectrum
anisotropy predictions.  The solid line shows the constraint on 
$\Omega_\Lambda$ as a function of the theoretical error on 
$\beta$, using $N_{pair}(<\theta) = 13 \left(\theta / 1'\right)^2$ 
full pairs.  The dashed line shows the same
thing for 10\% of the data.
}
\label{vardeltbeta}
\end{figure}

Throughout this subsection, I have been assuming that the simulation
predictions are essentially perfect, i.e., assuming that, given the 
input model parameters, I can compute the power spectrum exactly.
It is informative to relax this assumption, by allowing for an error,
$\Delta \beta$, in the predicted large-scale anisotropy parameter $\beta$.  
I treat $\beta$ as a new free parameter in the Fisher matrix 
calculation, and plot in Figure \ref{vardeltbeta} 
the resulting error in $\Omega_\Lambda$
as a function of the imposed constraint $\Delta \beta$.     
The solid line shows results using my usual 
set of pairs, and assuming only weak constraints on 
$\bF$, $T_{1.4}$, and $\gmo$.  
The precision of the $\Omega_\Lambda$ measurement
is reduced from $\Delta \Omega_\Lambda=0.037$ for perfectly predicted
$\beta$, to $\Delta \Omega_\Lambda = 0.059$ if $\beta$ is a completely 
free parameter.  The transition between the two cases actually occurs
at quite small values of $\Delta \beta / \beta$, because 
$\Delta \beta/ \beta$ can be measured to $0.074$ in a joint fit with 
$f(z)$.  This result is a target for simulators to shoot at --- 
predicting $\beta$ to $\sim 7$\% will allow a good consistency check,
and improving the prediction further will improve the measurement of
$\Omega_\Lambda$.  Finally, I show the results that can be obtained 
with only 10\% of the usual data, as the dashed line.  $\Omega_\Lambda$ 
can be constrained to almost $\pm 0.1$ if a good prediction for 
$\beta$ is available, and $\beta$ can be measured to $\sim 17$\% in
a joint fit.     

Unfortunately, as this paper was nearing completion, I discovered
that the SDSS magnitude limit for obtaining spectra at $z<3$ has
slipped to $i'< 19.1$ (X. Fan, private communication), which will
reduce the number of pairs by a factor of $\sim 5$ and correspondingly
increase the errors by $5^{1/2}$.  However, followup observations
of pair candidates should be able to recover the level of precision
I have discussed.  Reducing the limit by another magnitude 
(to $i'<20.5$) could increase the number of pairs by almost a factor    
of 10, giving a measurement of $\Omega_\Lambda$ to better than 2\%!

\section{CONCLUSIONS}

I have computed the three-dimensional power spectrum
of the transmitted flux, $P_F(\vk)$, using Hydro-PM (HPM) simulations.
From $P_F(\vk)$, we can calculate the correlation between 
absorption in the spectra of close pairs of quasars.     
The results for $P_F(\vk)$ for a set of model parameter variations 
are given in Table \ref{fitparms}, in terms of the parameters of 
equation (\ref{fitform}). 

I have investigated the importance of pressure, resolution, and box
size in the simulations.  Figure 
\ref{preseff} shows that the pressure force in the HPM simulations 
has an effect no larger than $\sim 10$\% on the 
large-scale power (relative to no pressure at all), 
but as much as $\sim 40$\% on small scales.  
However, within the HPM approximation, the change in pressure 
corresponding to a $\sim 4000$ K change in temperature
results in $\lesssim 2$\% change in power at all relevant $k$,
so it is not important to know the detailed thermal history
of the gas in order to compute the pressure.  
Figure \ref{restest}(a,b) shows that the required resolution for
convergence of the power (to $\sim 5$\%) on all relevant scales 
is $\sim 40\hkpc$,
while $80 \hkpc$ gives good results for 
$k \lesssim 1-2 \ihmpc$ only, and $160 \hkpc$ gives poor 
results on all scales.  Figure \ref{boxsize} shows that 
box size $L\sim 40 \hmpc$ is required for convergence of the
small-scale power, while reducing the box size to $L=20\hmpc$ 
leads to extra power along the line of sight.  My calculation
of the large-scale bias in \S 3.4 shows that $40 \hmpc$ simulations 
are sufficiently large for this purpose also. 
Future simulation work will be focused on determining the accuracy
of the HPM approximations, and running larger simulations.

One of the primary reasons to measure the correlation between
the \lyaf\ absorption in multiple lines of sight
is to determine the cosmological geometry through the 
Alcock \& Paczy\'nski (1979) test 
(McDonald \& Miralda-Escud\'e 1999; 
Hui, Stebbins, \& Burles 1999), which I show can break the 
degeneracy between matter and vacuum energy in a flat 
universe (see Figure \ref{ommomlfcont}), in a way that 
is {\it independent of the equation
of state of the vacuum energy} (Figure \ref{ommwfcont}).
I used my results for $P_F(\vk)$ to estimate the constraining power
of the AP test performed using future data sets.  
Figures \ref{basicsdssfcon} and \ref{vardeltbeta} show 
that, using $N_{pair}(<\theta)=13 \left(\theta/1'\right)^2$ fully
overlapping pairs of spectra at 
angular separation $< \theta$, with $\theta < 10'$,  
$\Omega_\Lambda$ can be measured
to between $\pm 0.03$ and $\pm 0.06$, depending on our ability to
accurately measure or calculate other parameters of the \lyaf\ model.
The Sloan Digital Sky Survey will only obtain spectra for a factor of 
$\sim 5$ fewer pairs than this, but followup observations of pair
candidates to magnitude $i'<19.5$ should be able to achieve the 
results 
discussed.  Figures \ref{lownoisesdssfcon} and \ref{highressdssfcon}
show that only small gains are obtainable by improving spectral quality 
beyond SDSS's signal-to-noise ratio, $S/N>10$ (for 1 \AA\ pixels), and 
resolution, 2 \AA\ (FWHM).     

The results in this paper should have a wide variety of applications.
For example, the power spectrum calculations should
be sufficiently accurate for comparison with a data set consisting of
all the currently known pairs (to be safe, $P_F(\vk)$ in this paper 
should not be trusted if better than 20\%
accuracy is important, although it may be more accurate than this).  
They will also be useful for planning
future observing programs.  The power of the \lyaf\ AP test in combination
with other measurements of cosmological parameters can be investigated.
Furthermore, other uses of the \lyaf\ to constrain cosmology can be 
explored using a realistic power spectrum,
e.g., using the very large quasar surveys to essentially measure $P_F(\vk)$
directly on very large scales (larger than the mean transverse quasar
separation), and constrain the detailed shape of the primordial
mass power spectrum [i.e., possibly measuring 
$\Gamma(z) \equiv \Omega_m h^2 (1+z) / H(z)$, or even detecting baryonic
wiggles].  
The computed bias parameters relating large-scale fluctuations in 
the \lyaf\ to fluctuations in the mass can be used to interpret 
correlations between the \lyaf\ and other observables, such as 
Lyman-break galaxies.
Finally, this paper should provide a useful starting point
for future simulation projects that seek to compute $P_F(\vk)$ more 
accurately.  

\acknowledgements

I thank David Weinberg, Uros Seljak, David Tytler,  
Andy Albrecht, and especially Jordi Miralda-Escud\'e for helpful 
conversations and/or comments on the manuscript, and Nick Gnedin 
for his HPM code.

\begin{deluxetable}{lcccccccccc}
\tablecolumns{11}
\tablecaption{Power Spectrum Results\label{fitparms}}
\tablehead{
\colhead{$\delta {\mathbf p}$} & \colhead{$b_\delta^2$} & \colhead{$\beta$} & 
\colhead{$k_{NL}$} & \colhead{$\alpha_{NL}$} & \colhead{$k_P$} & 
\colhead{$\alpha_P$} & \colhead{$k_{V0}$} & \colhead{$\alpha_V$} & \colhead{$k'_V$} & \colhead{$\alpha'_V$}} 
\startdata
0 & 0.0173 & 1.58 & 6.40 & 0.569 & 15.3 & 2.01 & 1.220 & 1.50 & 0.923 & 0.451 \\
   $+\delta A_1$   & 0.0151 & 1.66 & 6.59 & 0.524 & 16.0 & 2.11 & 0.963 & 1.50 & 0.727 & 0.487 \\
   $-\delta A_1$   & 0.0197 & 1.51 & 6.30 & 0.568 & 15.2 & 1.94 & 1.430 & 1.50 & 0.843 & 0.391 \\
$+\delta n_1$   & 0.0171 & 1.62 & 8.30 & 0.551 & 16.0 & 2.05 & 1.183 & 1.48 & 0.920 & 0.447 \\
$-\delta n_1$   & 0.0175 & 1.54 & 5.07 & 0.579 & 14.5 & 1.94 & 1.254 & 1.53 & 0.929 & 0.456 \\
 $+\delta T_{1.4}$   & 0.0173 & 1.58 & 6.41 & 0.571 & 15.2 & 2.01 & 1.205 & 1.51 & 0.829 & 0.431 \\
   $-\delta T_{1.4}$  & 0.0173 & 1.58 & 6.42 & 0.570 & 15.3 & 2.00 & 1.234 & 1.50 & 1.025 & 0.473 \\
$+\delta(\gmo)$ & 0.0171 & 1.57 & 6.42 & 0.568 & 15.2 & 2.03 & 1.242 & 1.48 & 0.959 & 0.443 \\
$-\delta(\gmo)$ & 0.0176 & 1.59 & 6.39 & 0.576 & 15.2 & 1.97 & 1.196 & 1.53 & 0.885 & 0.459 \\
$+\delta \bF$ & 0.0126 & 1.49 & 5.21 & 0.707 & 13.1 & 1.83 & 1.476 & 1.43 & 2.135 & 0.469 \\
$-\delta \bF$& 0.0235 & 1.66 & 5.62 & 0.613 & 12.6 & 1.70 & 0.895 & 1.54 & 0.431 & 0.464 \\
\enddata
\tablecomments{$k$'s measured in $\ihmpc$.  $\delta A_1=0.29$, $\delta n_1=0.1$
(at fixed $\sigma_1$), $\delta T_{1.4}=2000$ K, $\delta (\gmo)=0.05$, and
$\delta \bF = 0.05$.}
\end{deluxetable}

\end{document}